\documentclass[twocolumn,trackchanges]{aastex63}

\shorttitle{LAMOST-MRS Spectroscopic candidates}
\shortauthors{LI et al.}

\begin{document}

\title{Double-, triple-line spectroscopic candidates in the LAMOST Medium-Resolution Spectroscopic Survey}

\author[0000-0002-6647-3957]{Chun-qian Li}
\affiliation{CAS Key Laboratory of Optical Astronomy, National Astronomical Observatories, Chinese Academy of Sciences, Beijing 100101, China}
\affiliation{School of Astronomy and Space Science, University of Chinese Academy of Sciences, Beijing 100049, China}

\author[0000-0002-0349-7839]{Jian-rong Shi}
\affiliation{CAS Key Laboratory of Optical Astronomy, National Astronomical Observatories, Chinese Academy of Sciences, Beijing 100101, China}
\affiliation{School of Astronomy and Space Science, University of Chinese Academy of Sciences, Beijing 100049, China}
\email{sjr@nao.cas.cn}

\author[0000-0002-8609-3599]{Hong-liang Yan}
\affiliation{CAS Key Laboratory of Optical Astronomy, National Astronomical Observatories, Chinese Academy of Sciences, Beijing 100101, China}
\affiliation{School of Astronomy and Space Science, University of Chinese Academy of Sciences, Beijing 100049, China}
\email{hlyan@nao.cas.cn}

\author[0000-0001-8241-1740]{Jian-Ning Fu}
\affiliation{Department of Astronomy, Beijing Normal University, Beijing 100875, P.R.China}

\author[0000-0002-3651-5482]{Jia-dong Li}
\affiliation{CAS Key Laboratory of Optical Astronomy, National Astronomical Observatories, Chinese Academy of Sciences, Beijing 100101, China}
\affiliation{School of Astronomy and Space Science, University of Chinese Academy of Sciences, Beijing 100049, China}

\author[0000-0002-3701-6626]{Yong-Hui Hou}
\affiliation{School of Astronomy and Space Science, University of Chinese Academy of Sciences, Beijing 100049, China}
\affiliation{Nanjing Institute of Astronomical Optics \& Technology, National Astronomical Observatories, Chinese Academy of Sciences, Nanjing 210042, China}

\begin{abstract}

The LAMOST Medium-Resolution Spectroscopic Survey (LAMOST-MRS) provides an unprecedented opportunity for detecting multi-line spectroscopic systems. Based on the method of Cross-Correlation Function (CCF) and successive derivatives, we search for spectroscopic binaries and triples and derive their radial velocities (RVs) from the LAMOST-MRS spectra. A Monte-Carlo simulation is adopted to estimate the RV uncertainties. After examining over 1.3 million LAMOST DR7 MRS blue arm spectra, we obtain 3,133 spectroscopic binary (SB) and 132 spectroscopic triple (ST) candidates, which account for 1.2\% of the LAMOST-MRS stars. Over 95\% of the candidates are newly discovered. It is found that all of the ST candidates are on the main sequence, while around 10\% of the SB candidates may have one or two components on the red giant branch.

\end{abstract}
\keywords{Catalogs -- Spectroscopic binary stars -- Radial velocity}

\section{Introduction}\label{sec: intro}

Since approximately half of the stars in our Galaxy are in double, triple or high-order systems \citep{Raghavan_2010}, multiple star systems play an essential role in astrophysics, especially binary systems. The characterization of multiple star systems such as orbital parameters can be investigated by combining the spectroscopic and photometric information.

Spectroscopic multiple star systems can be classified according to the number of stellar components in the spectra, because their spectral lines split due to the different radial velocities (RVs). A target can be considered as a spectroscopic binary (SB) or spectroscopic triple (ST) candidate if it has a double- or triple-line spectrum. The single-line SBs can also be identified as their radial velocities are variable. 

Up to now, some catalogs of the spectroscopic multiple star systems have been published, $\rm S_{B^9}$, the recent version of the ninth catalogue of spectroscopic binary orbits includes more than 4,004 SBs, and around one third of them are double-line systems \citep{SB9}, and the Geneva-Copenhagen Survey Catalogue presents 3,223 SBs from 16,682 nearby F and G dwarf stars \citep{GCS2004, GCS2009}. The recent spectroscopic surveys provide a bulk of spectra, which have largely expanded the number of spectroscopic multiple star systems, such as the Radial Velocity Experiment (RAVE) survey \citep{RAVE_SB2} has found 123 SBs, the \textit{Gaia}-ESO survey \citep{Gaia-ESO_2017} has detected 342 SB, 11 ST and even 1 quadruple-line candidates, the Apache Point Observatory Galaxy Evolution Experiment (APOGEE) survey has discovered more than 3,000 binaries \citep{APOGEE/IN-SYNC_2017, APOGEE_Bianry_2018}, and the Galactic Archaeology with HERMES (GALAH) survey has derived 12,760 FGK SBs with stellar properties \citep{GALAH_binary}. 

The Large Sky Area Multi-Object fiber Spectroscopic Telescope (LAMOST) is a 4-meter Schmidt telescope with a 5\degr\ field of view (FOV), and is equipped with 4,000 fibers \citep{LAMOST_2012, LAMOST_DR1, Zhao2006, Zhao2012}. The medium-resolution spectroscopic (MRS) survey includes both blue and red arms, and the wavelength coverages of them are from 4,950 to 5,350\,\AA\, and from 6,300 to 6,800\,\AA, respectively. The resolution power of MRS is 7,500. There are more absorption lines in the blue arm than those in the red one, which enables us to measure the RVs with a precision of 1\,km/s for most of the stars \citep{Liu2019}. The efficiency of observations and the achieved precisions make LAMOST-MRS an exceptional database for measuring RVs and detecting multiple star systems.

The purpose of this work is to detect the multiple-line candidates (SB and ST) from the LAMOST-MRS database following the method of \citet{Gaia-ESO_2017}. This paper is organized as follows, in Sect.\,\ref{sec: data}, we describe the LAMOST-MRS spectral dataset. The methods used to normalize spectra, calculate CCFs and detect peak positions in CCFs are introduced in Sect.\,\ref{sec: method}, and a catalog that contains SB and ST candidates detected from LAMOST-MRS is presented in Sect.\,\ref{sec: result}. In Sect.\,\ref{sec: discuss}, we discuss the detection efficiency, the stellar parameters of the SB and ST candidates, radial velocity differences and caveats, and the conclusions are given in Sect\,\ref{sec: conclusion}.

\section{Data Selection} \label{sec: data}

\begin{figure}[!t]
    \begin{minipage}{0.5\linewidth}
        \centering
        \includegraphics[width = 3in]{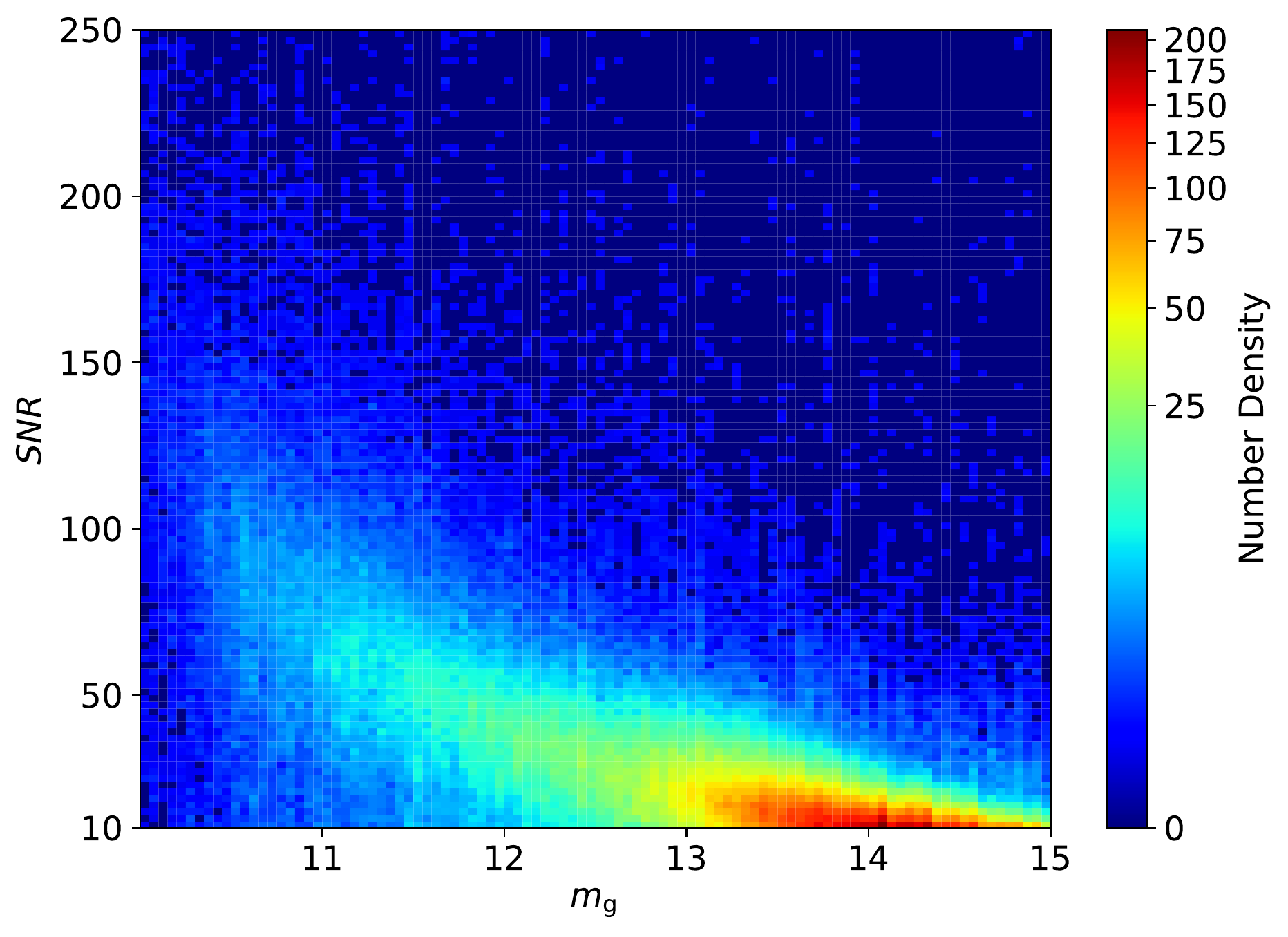}
    \end{minipage} \\
    \begin{minipage}{0.5\linewidth}
        \centering
        \includegraphics[width = 3in]{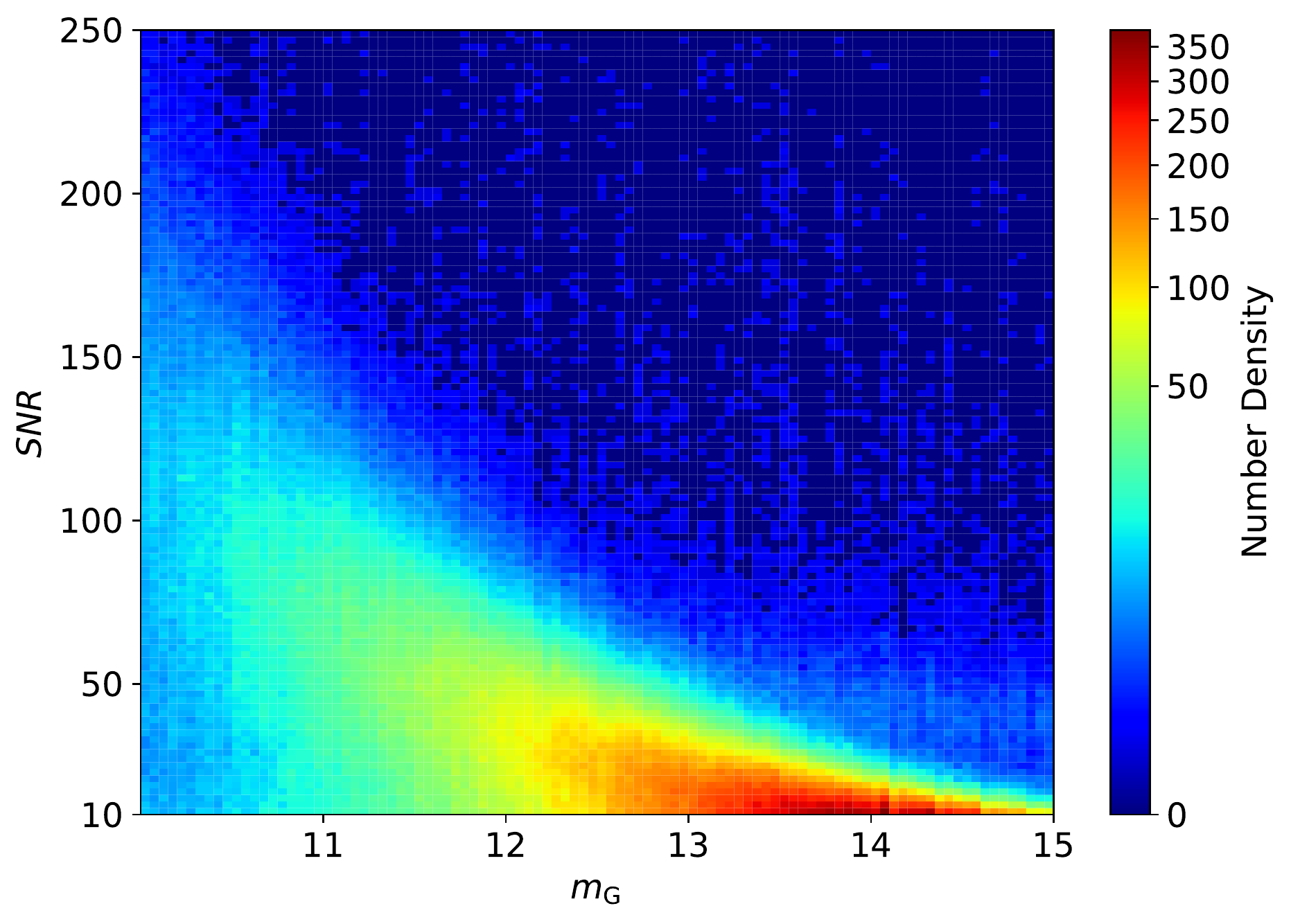}
    \end{minipage}
    \caption{The distributions of SNR versus SDSS \textit{g} magnitude of the first year survey (upper panel) and the \textit{Gaia} DR2 G magnitude of the second year survey (lower panel).
    \label{fig: LAMOST_mag_snr}}
\end{figure}

The LAMOST-MRS test observation started in Sep. 2017, and the LAMOST-MRS survey began in Oct. 2018. The first and second year MRS data has been released in the LAMOST Data Release 7\footnote{\url{http://dr7.lamost.org/}} (DR7). The wavelength ranges of the LAMOST-MRS blue and red arms are from 4,950 to 5,350\,\AA\ and from 6,300 to 6,800\,\AA. Until April 11, 2019, the LAMOST-MRS survey has obtained 5,369,891 spectra of 759,886 objects. Only the spectra of blue arms have been chosen to detect multiple line candidates and measure their RVs, as there are more absorption lines in the blue arms. The distributions of the signal-to-noise ratio (SNR) of the blue arm spectra versus the SDSS \textit{g} magnitude of the first year survey and the \textit{Gaia} DR2 G magnitude of the second year survey are displayed in Figure\,\ref{fig: LAMOST_mag_snr}. It can be seen that most of the objects have \textit{g} or G magnitudes between 10 and 15. It is found that a spectrum with low SNR will lead to challenges for detecting reliable SBs and STs \citep{APOGEE/IN-SYNC_2017}. Therefore, the blue arm spectra with SNR higher than 10 have been selected. In addition to SNR, the header keywords \textit{objtype} and \textit{fibermas} have also been taken into consideration. Only these spectra with \textit{objtype} of star and \textit{fibermas} of 0 have been chosen, because the header keyword \textit{fibermas} of not 0 indicates that this fiber may have problems. Finally, 1,383,831 spectra of 281,437 stars have been selected.

\begin{figure}[!t]
    \centering
    \plotone{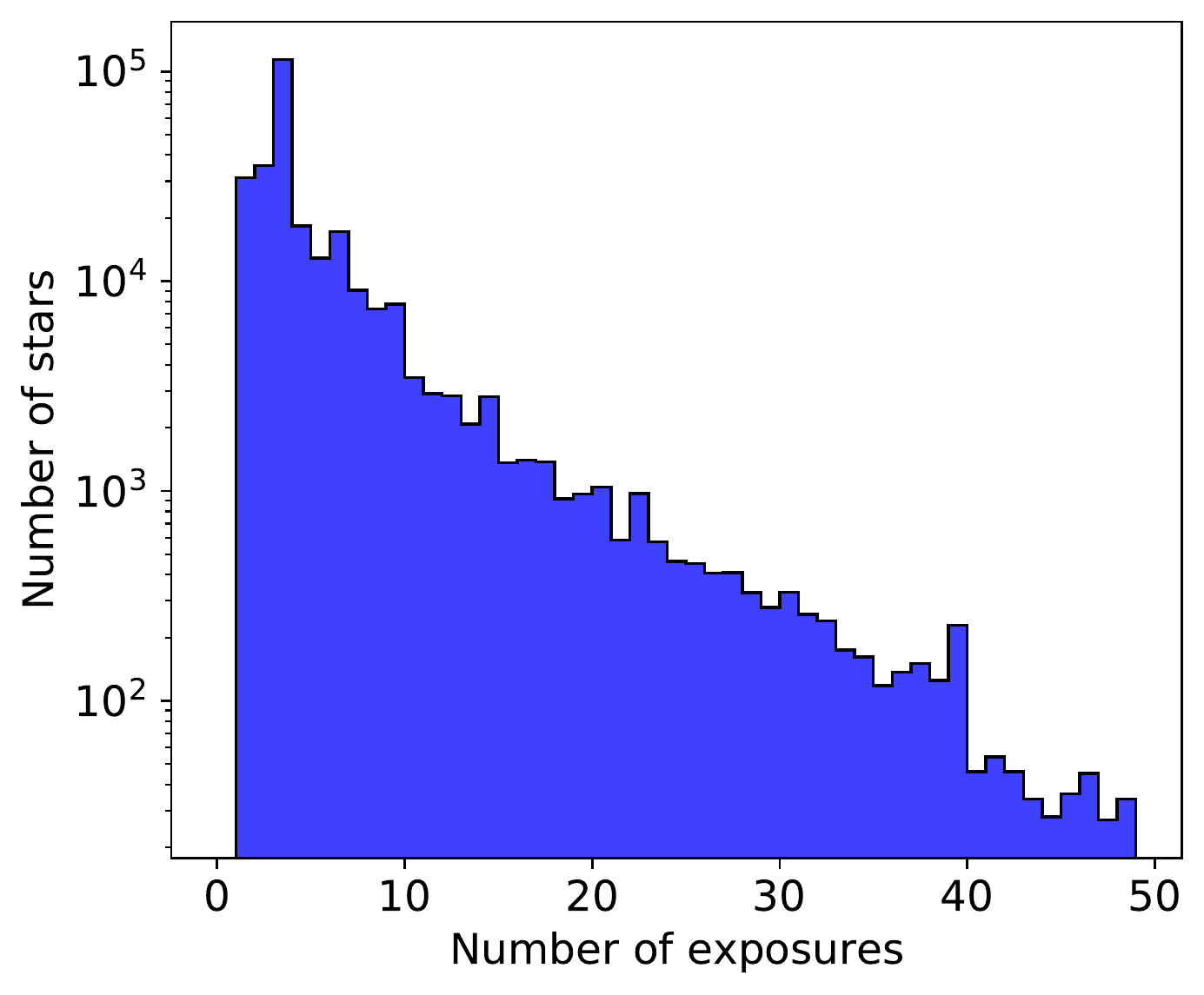}
    \caption{Number of selected stars versus number of exposures. The number of stars is in {$log$} scale.}
    \label{fig: star_epoch}
\end{figure}

\begin{figure}[!t]
    \plotone{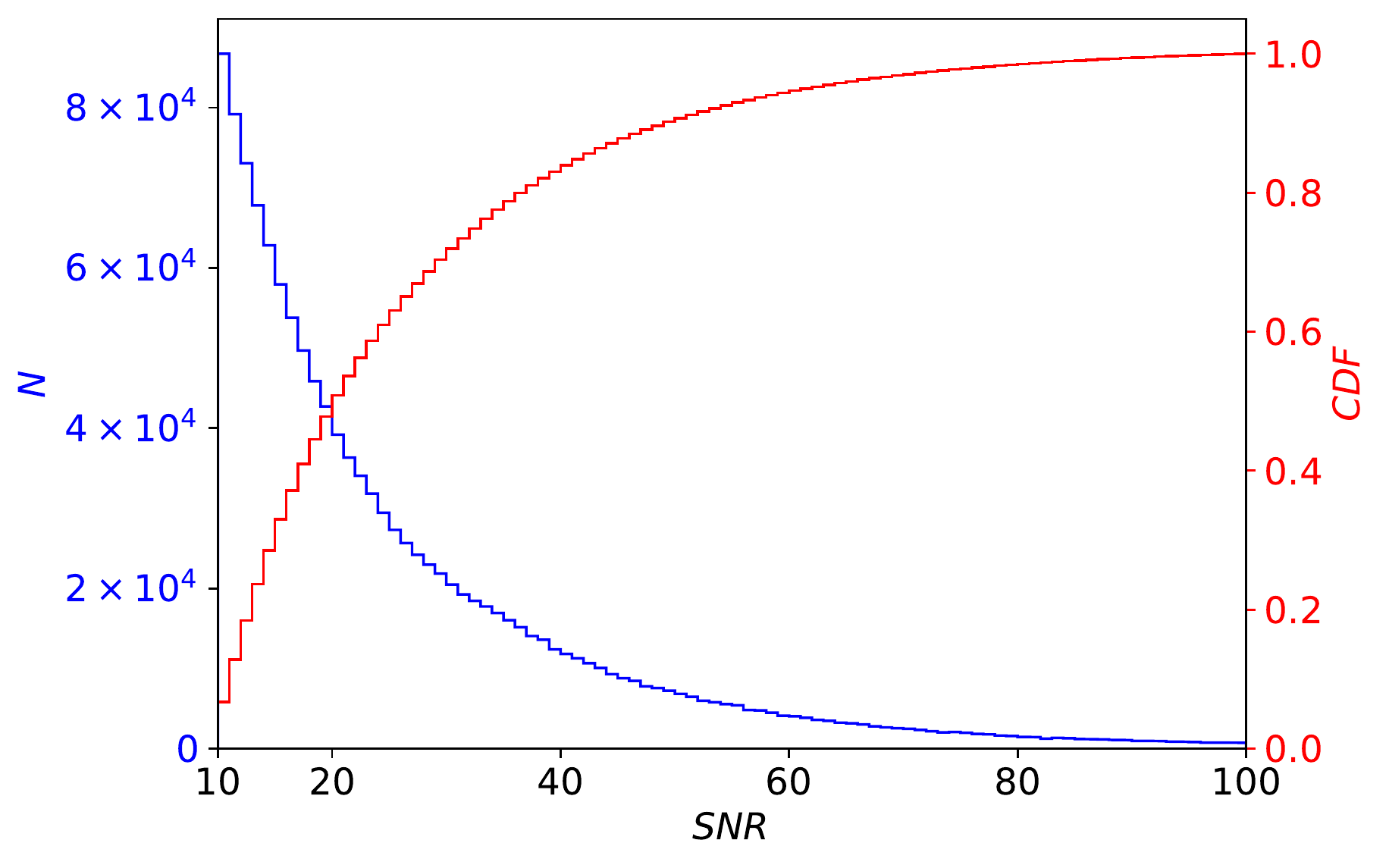}
    \caption{Number and Cumulative Distribution Function (CDF) diagram of SNR of all the selected spectra.}
    \label{fig: SNR_CDF}
\end{figure}

Figure\,\ref{fig: star_epoch} shows the number of exposures for these stars, and most of them have been observed three times because the LAMOST-MRS no-time-domain observation takes three consecutive exposures \citep{Liu2020}. The number and Cumulative Distribution Function (CDF) versus SNR of all the selected spectra are presented in Figure\,\ref{fig: SNR_CDF}.

\section{Method} \label{sec: method}

Based on the \textit{Gaia}-ESO survey, \citet{Gaia-ESO_2017} have developed a method to search for multi-line spectroscopic candidates. Their method utilizes the Gaussian kernel to smooth the successive derivatives of Cross-Correlation Function (CCF). Following their method, we detect SB and ST candidates and derive their RVs from the LAMOST-MRS spectra. A Monte-Carlo (MC) simulation is adopted to estimate the RV uncertainties for each spectrum.

\subsection{Calculating CCF} \label{subsec: method_CCF}

\begin{figure*}[!t]
    \centering
    \includegraphics[width = 6in]{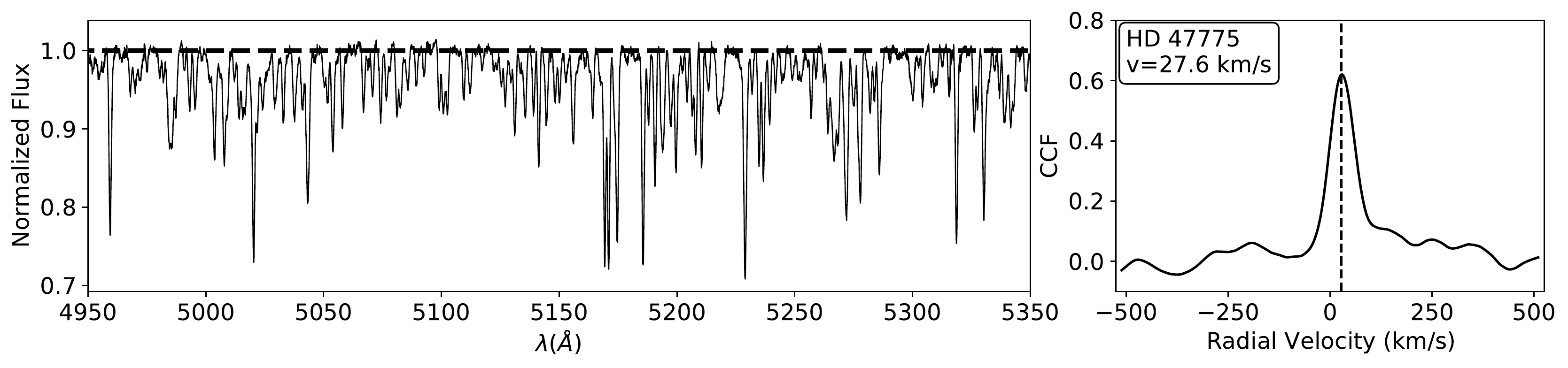}
    \caption{The normalized LAMOST-MRS spectra (\textit{left panel}) and CCF diagrams (\textit{right panel}) of HD\,47775.}
    \label{fig: spec_CCF}  
\end{figure*}

CCF is usually used to measure RVs and to find multiple line spectra. We use Eq.\,\ref{eq: normalized_CCF} to calculate the values of the normalized CCF according to the normalized correlation coefficient \cite[p.~92]{gubner2006probability}. The range of the normalized CCFs is between $-$1 and $+$1, with $+$1 indicating a perfect correlation and $-$1 for a perfect anti-correlation.

\begin{equation} \label{eq: normalized_CCF}
    CCF(v) = \displaystyle \sum_{i=1}^n\frac{(O_i - \overline O)}{\sigma_O}\cdot\frac {(T_{i,v} - \overline{T})}{\sigma_T},
\end{equation}

\noindent Here, $O_i$ and $T_{i,v}$ are the normalized flux of the observed and template spectra at a same wavelength sampling point $i$, and the RV value of the template is located at \textit{v}. $\overline{O}$ and $\sigma_O$ are the mean and scatter of the flux values of the observed spectrum, while $\overline{T}$ and $\sigma_T$ for the template. In this work, we generate a set of template spectra with a series of RVs based on the normalized solar spectrum \citep{KPNO1984}. The RV variations are from $-$500 to $+$500\,km/s with a step of 1\,km/s. To decrease the sampling points, we reduce the resolution of the templates to 100,000.

It is noted that the peaks in some CCFs are too low to be detected, therefore, these spectra with a maximum value of CCF less than 0.2 and/or the difference between the maximum and minimum values of CCF less than 0.1, have been not considered. About 80,000 spectra have been excluded by these criteria.

The spectra need to be normalized, and we apply a general normalization procedure to all the selected spectra. The normalization procedure is as follows:

\begin{enumerate}
    \item Splitting each spectrum evenly into 10 bins by wavelength,
    \item Quadratic spline interpolation with the median flux values of the 10 bins,
    \item Masking the cosmic-rays, absorption and emission lines.
\end{enumerate}

As the cosmic-rays and emission lines will lead to negative values of CCF, they need to be masked during normalization. Thus, we calculate the standard deviation ($\sigma$) of the residual between the observed spectrum and the continuum, and mask these points where the residuals are higher than $+5\sigma$.

The whole process is iterated three times to improve the continuum. The left panel of Figure\,\ref{fig: spec_CCF} shows an example, the normalized spectrum of HD\,47775.

\subsection{Detecting the number of RV components and determining their RVs} \label{subsec: method_DetPeaks}

\begin{figure*}[!t]
    \centering
    \includegraphics[width = 2.3in]{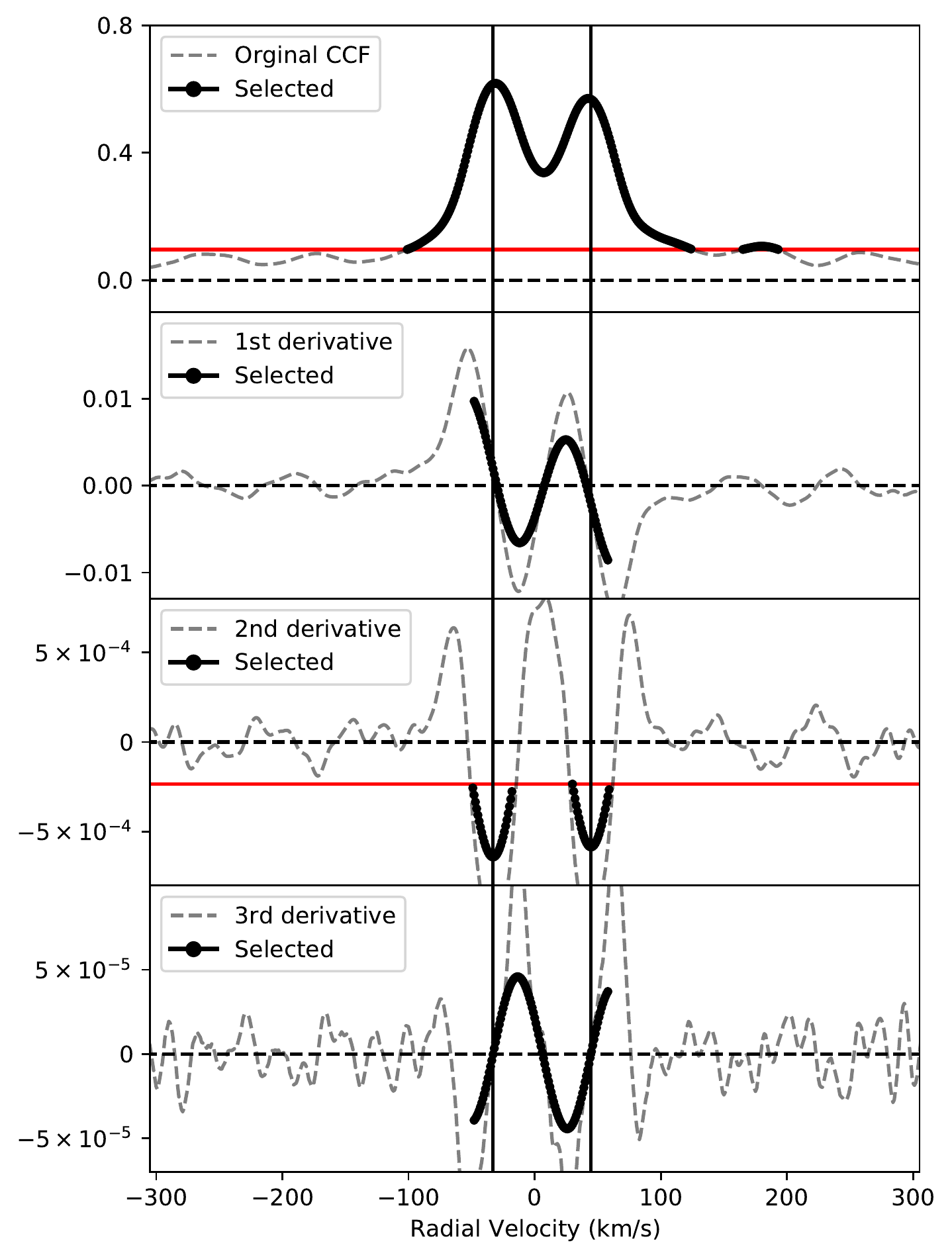}
    \includegraphics[width = 2.3in]{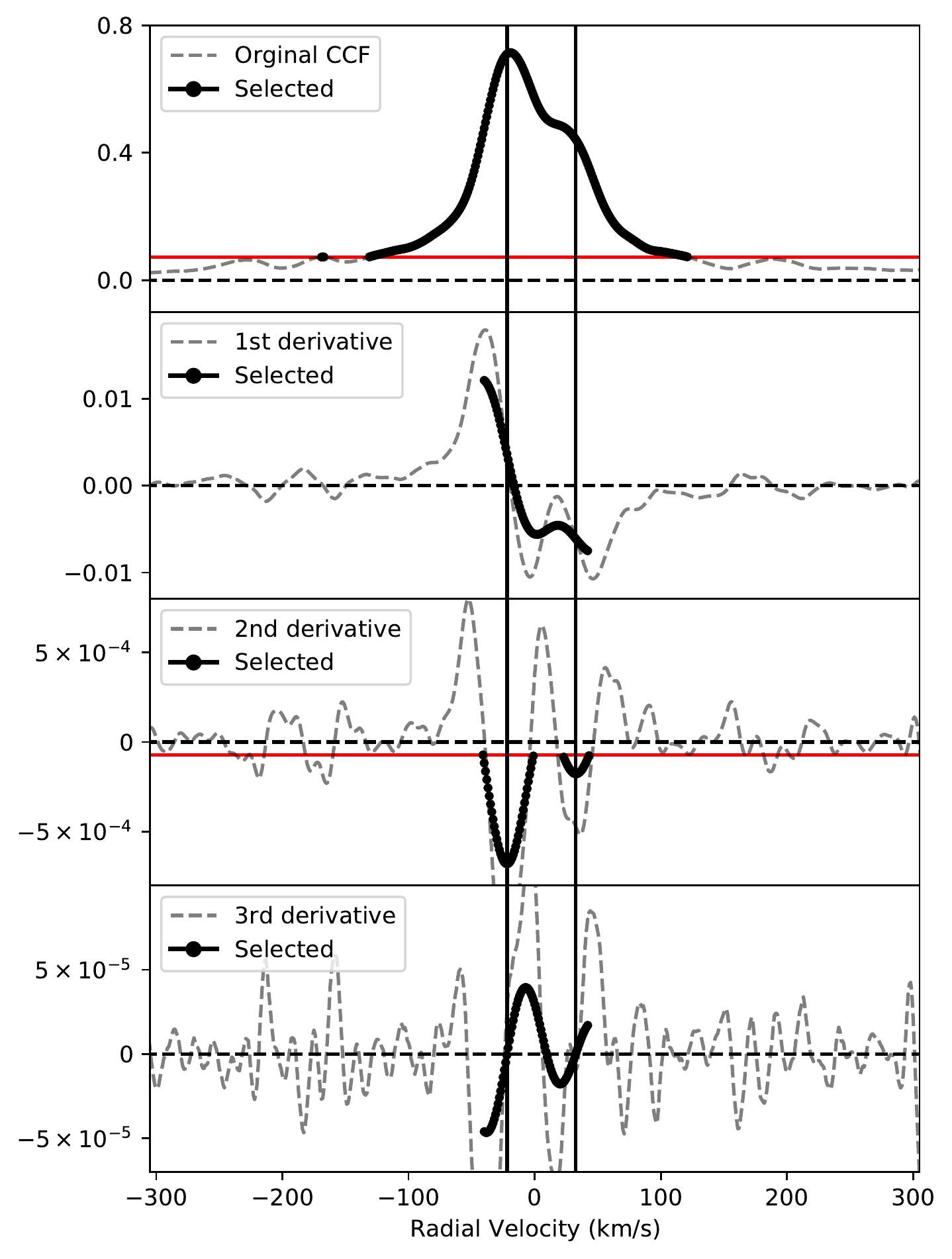}
    \includegraphics[width = 2.3in]{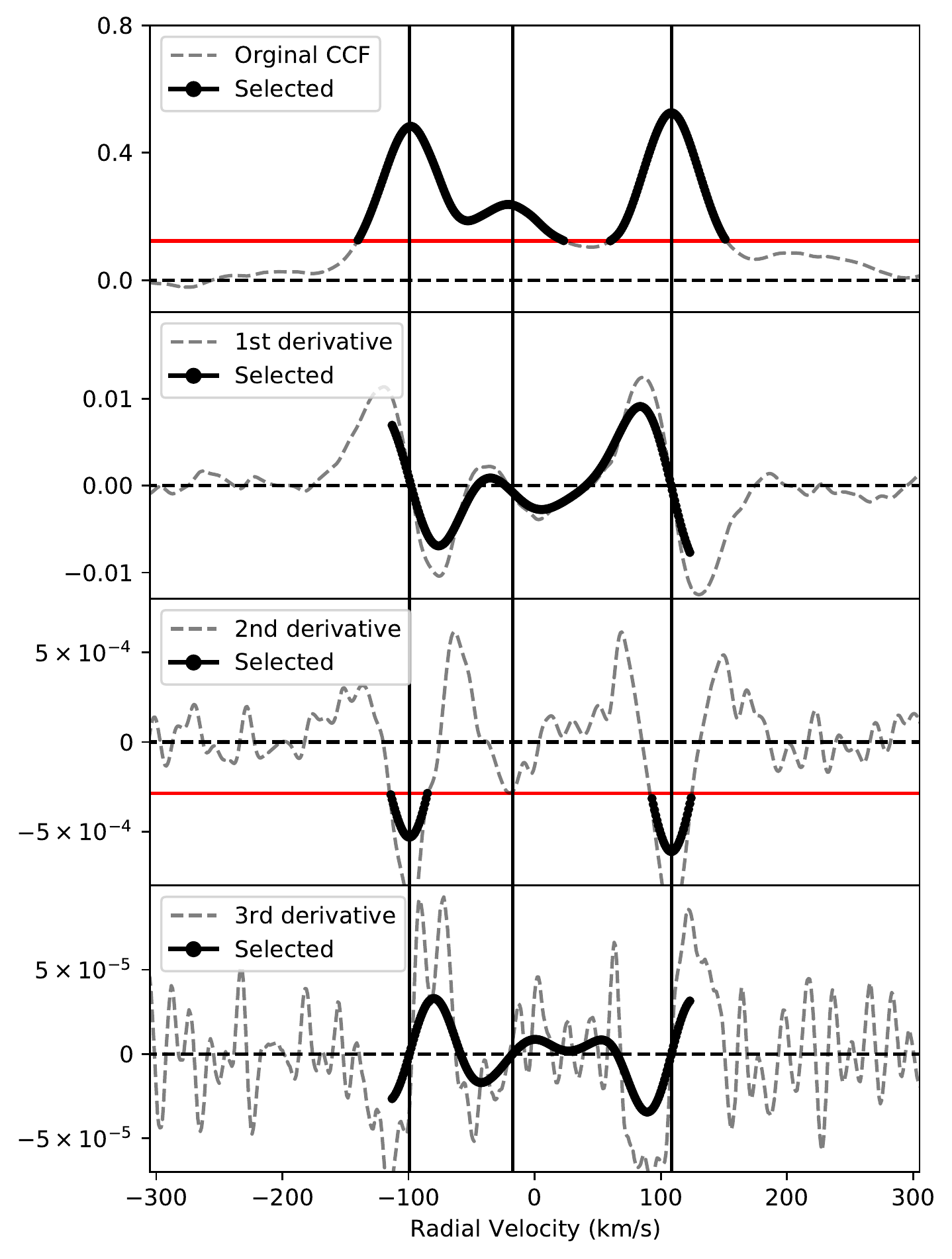}
    \caption{CCFs and derivatives of two SB and one ST candidates. The gray dashed lines are the original CCFs and derivatives, all the selected range of CCFs and smoothed derivatives are drawn with black solid lines. The red horizontal lines are the thresholds for each candidate, and the black vertical lines are their RVs.}
    \label{fig: Blend_peaks}
\end{figure*}

\begin{figure*}[!t]
    \centering
    \includegraphics[width = 3in]{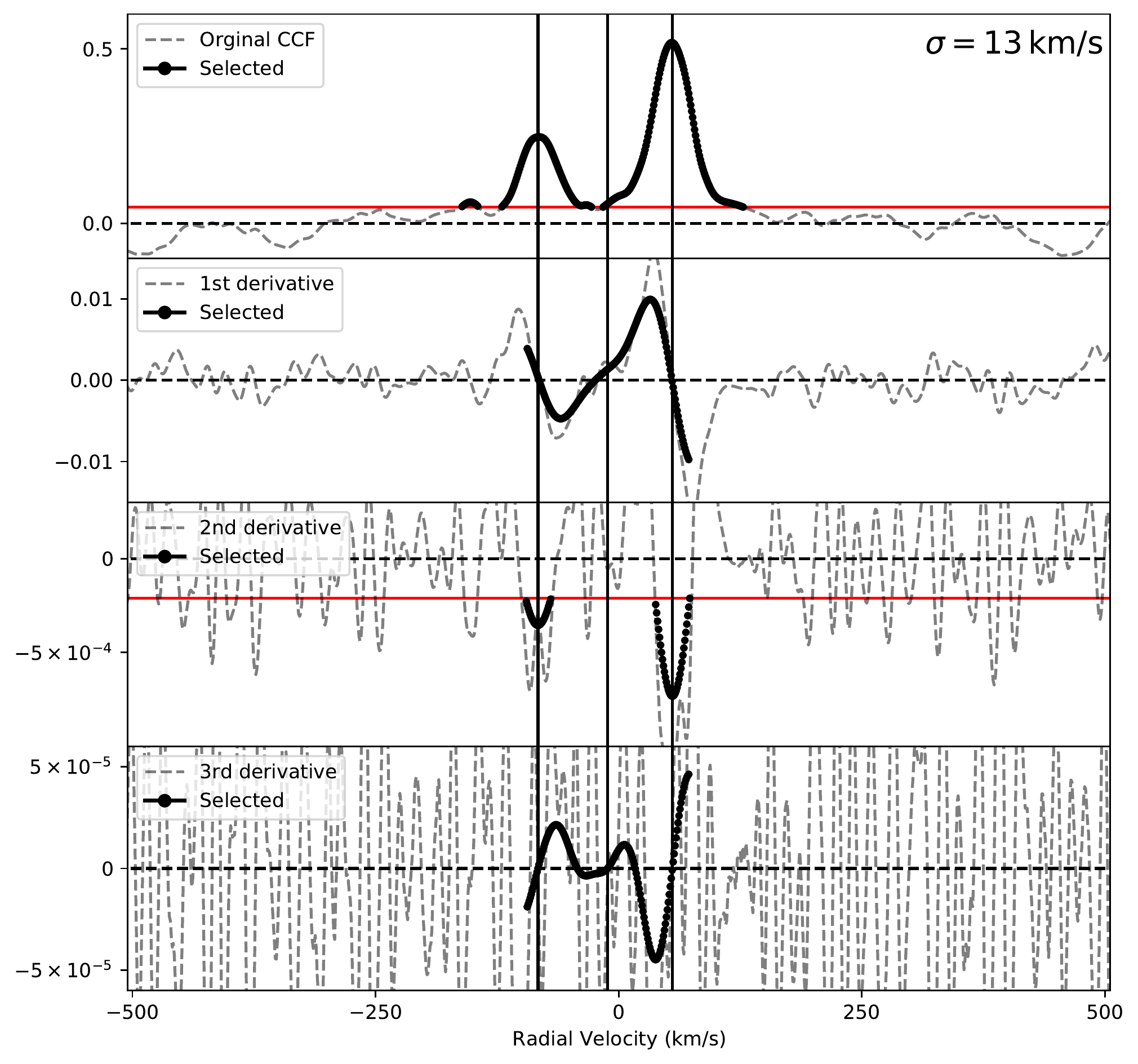}
    \includegraphics[width = 3in]{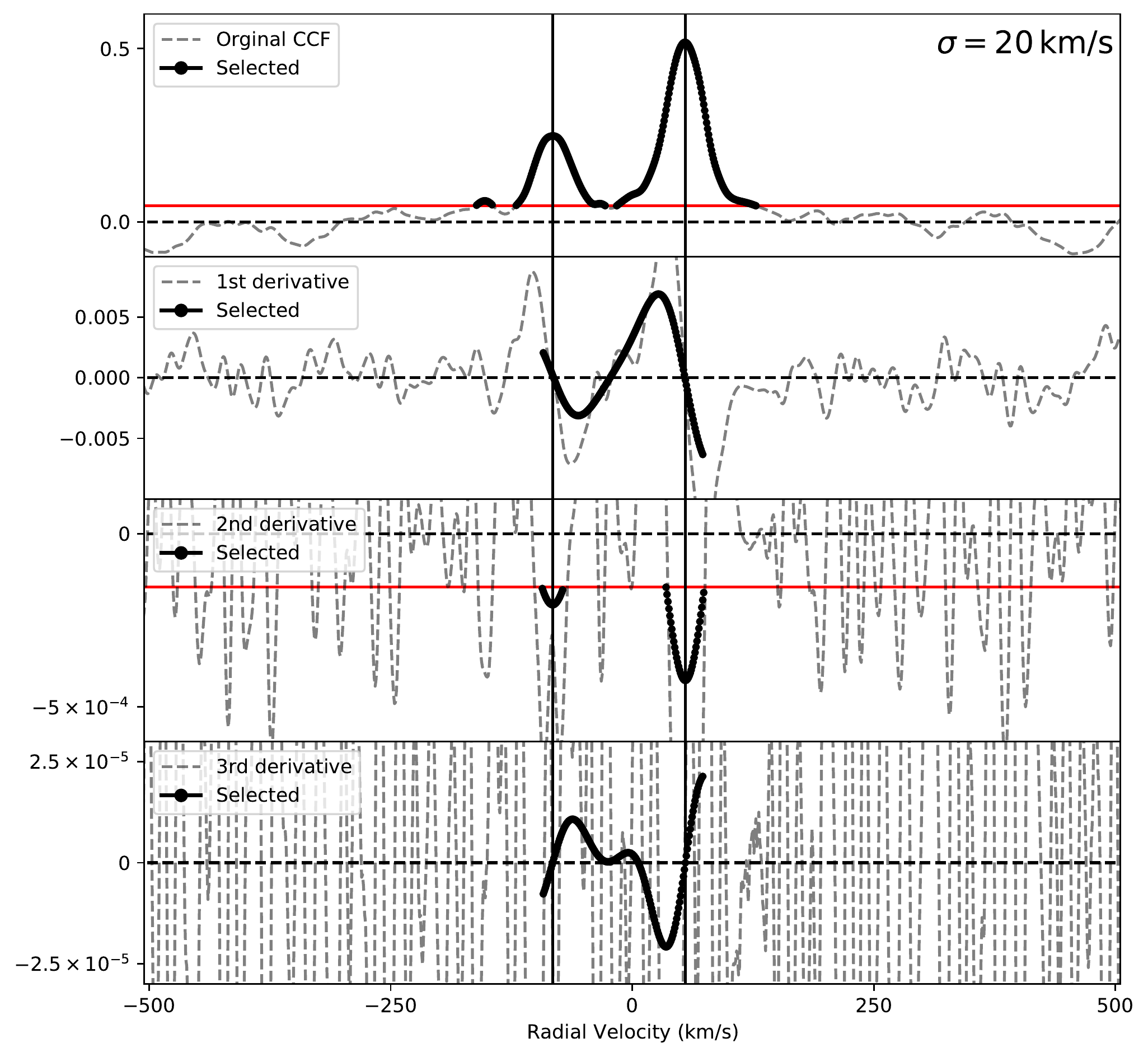}
    \caption{The example of finding an appropriate $\sigma$ of Gaussian smoothing.}
    \label{fig: Gauss_sigma}
\end{figure*}

To obtain the number of RV components and their values, \cite{Gaia-ESO_2017} have designed a detection of extrema code to identify the peaks in CCFs. In their Figure\,1, \citet{Gaia-ESO_2017} presented simulated CCFs and their derivatives, there is a peak in the CCF diagram, where is the value of RV. By detecting where the 1st derivative passes zero in declining phase or the 3rd derivative passes zero in ascending phase, it is possible to identify the RV components of the spectrum and calculate their RV values.

We note that small bulges (CCF values are around zero) can also be identified as peaks for the 1st or 3rd derivatives, for instance, in the right panel of Figure\,\ref{fig: Blend_peaks}. The reason is that there are more noises in the higher-order derivatives for a discrete CCF. To avoid the impact of these spurious peaks, it is necessary to select the RV range and smooth the derivatives. Therefore, the Gaussian filter function \textit{gaussian\_filter1d} of the \textit{scipy.ndimage} package \citep{SciPy} in Python is adopted to smooth the derivatives.

Two \textit{thresholds} are used to select the RV range, the first one is the percentile of the CCF, and the other one is the percentile of the smoothed 2nd derivative. Only the ranges where the CCF values are higher than the 75th percentile and the smoothed 2nd derivative values are lower than the 6th percentile have been selected, as shown in Figure\,\ref{fig: Blend_peaks}. If the selected ranges are separate, some triple-line spectra would be identified as double-line (see the right panel of Figure\,\ref{fig: Blend_peaks}). Thus, the middle part between the separate ranges are also selected.

\begin{figure}[!t]
    \plotone{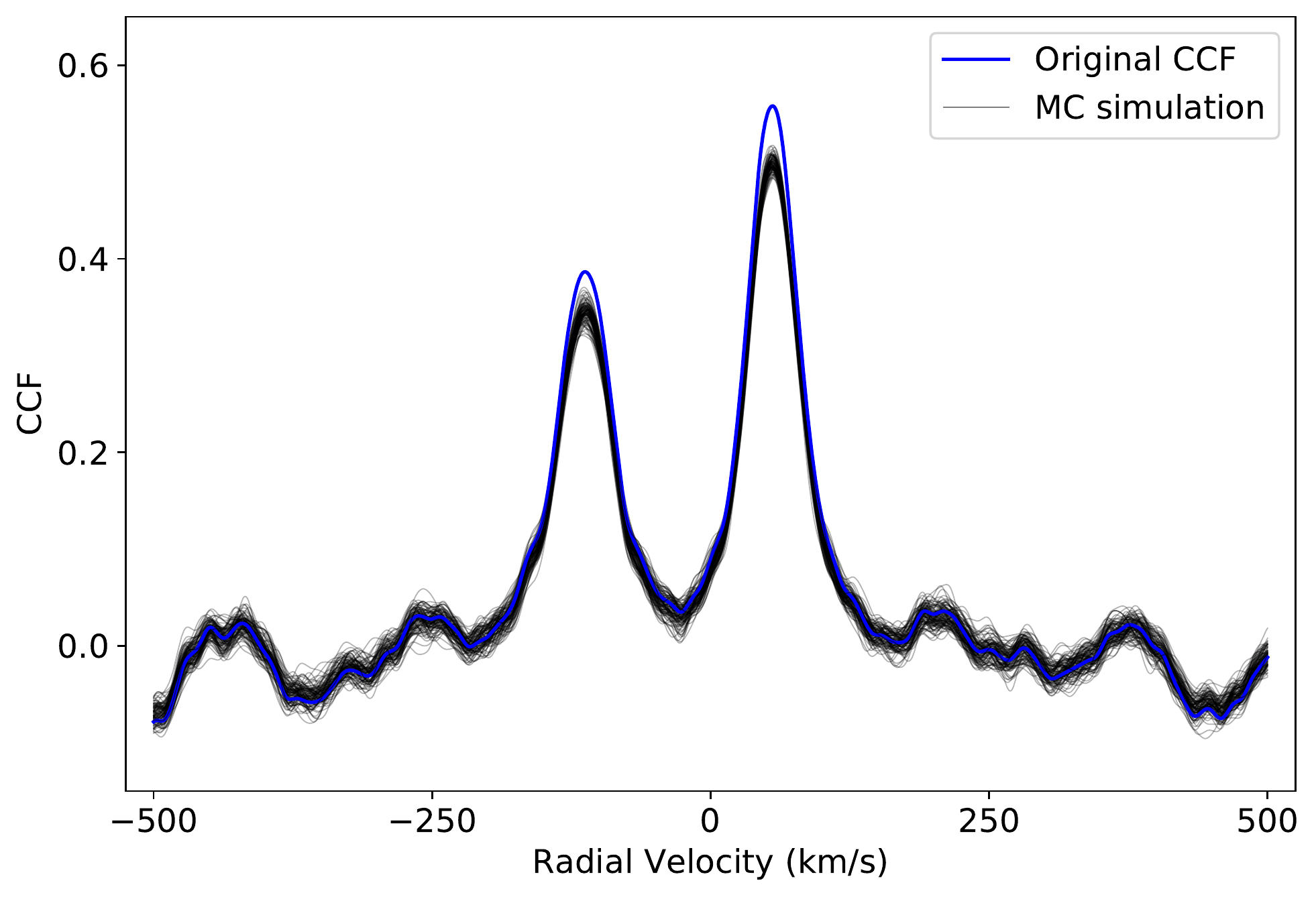}
    \caption{Original CCF of a double-line spectrum and MC simulation ones.}
    \label{fig: MC_CCF}
\end{figure} 

\begin{figure}[]
    \plotone{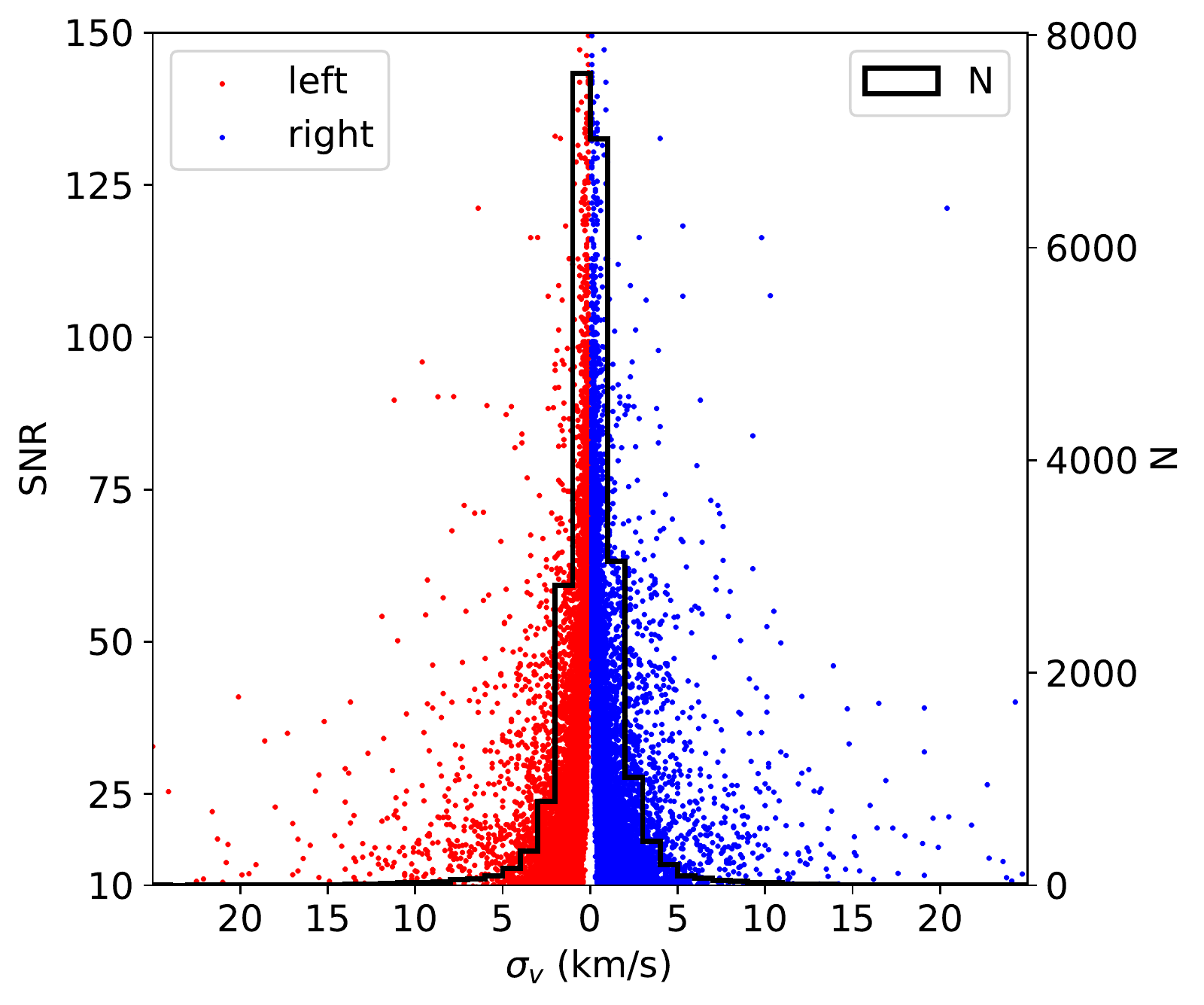}
    \caption{The SNR of spectra verses the RV uncertainties of SB candidates. Red and blue dots are the radial velocity of the left and right peak of CCF.}
    \label{fig: SNR_dv}
\end{figure}

As shown in the middle panel of Figure\,\ref{fig: Blend_peaks}, the two peaks are strongly blended in the CCF diagram, and only one peak can be detected by the 1st derivative. While, there are still two peaks can be detected from the 3rd derivative, therefore, we derive the RVs from the 3rd derivatives. A linear fit using the two near zero points of the ascending 3rd derivative is adopted to calculate the RV value.

An appropriate standard deviation ($\sigma$) of the Gaussian kernel is important for smoothing, and we choose an initial $\sigma$=13\,km/s for the LAMOST-MRS spectra. However, this value is too small for some cases, and it may lead to extra spurious RV component as shown in the left panel of Figure\,\ref{fig: Gauss_sigma}. In this case, we increase the $\sigma$ by 1\,km/s till the number of RV components detected with the 3rd derivative is equal to the number of valleys in the 2nd derivative, and the $\sigma$ has been increased from 13\,km/s to 20\,km/s (see the right panel of Figure\,\ref{fig: Gauss_sigma}).

\subsection{Uncertainties} \label{subsec: result_eRV}

\begin{figure}[!t]
    \plotone{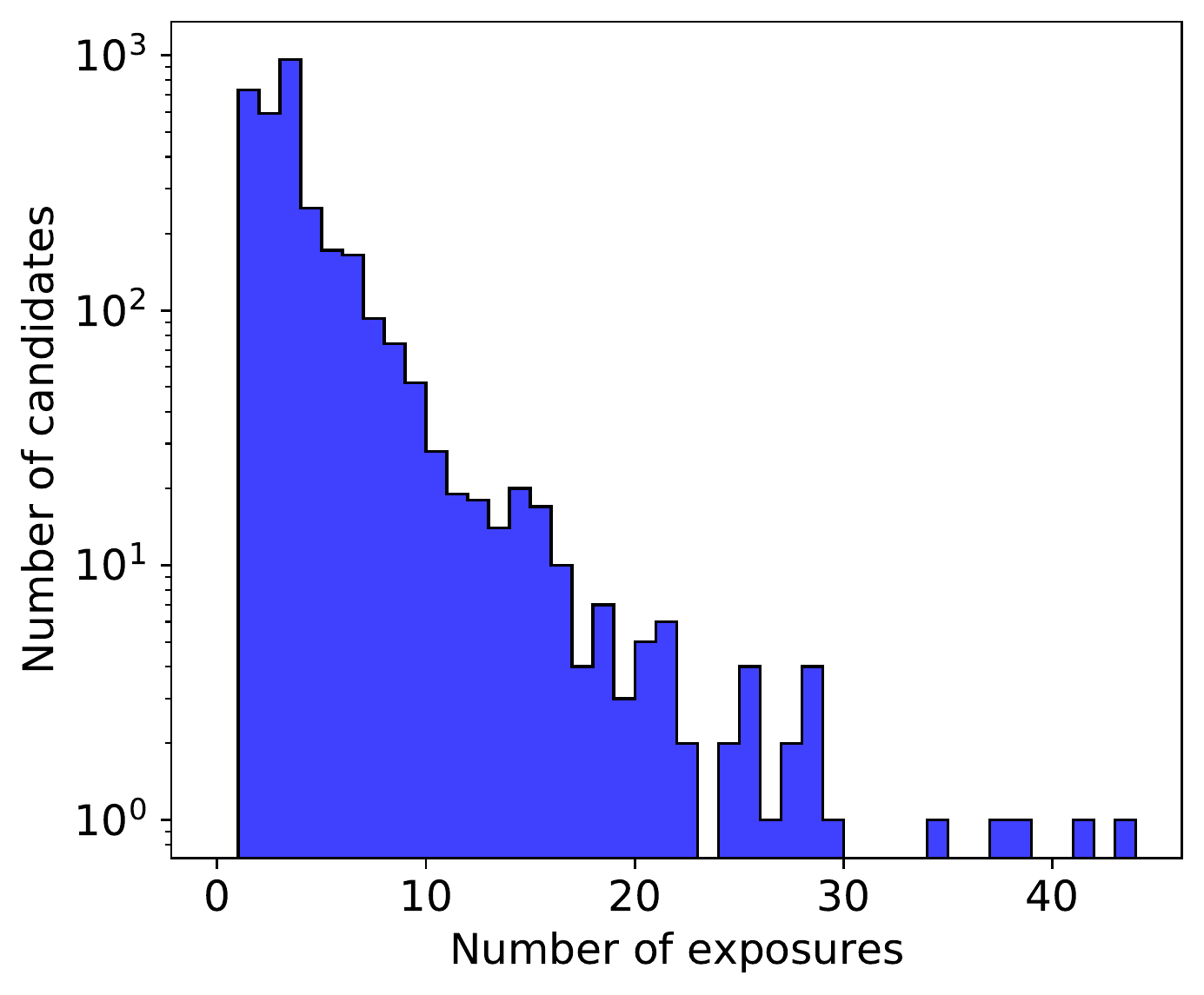}
    \caption{Number of the detected SB and ST candidates versus number of exposures. The number of stars is in {$log$} scale.}
    \label{fig: SB_ST_epoch}
\end{figure} 

\begin{figure*}[!t]
    \centering
    \includegraphics[width = 3in]{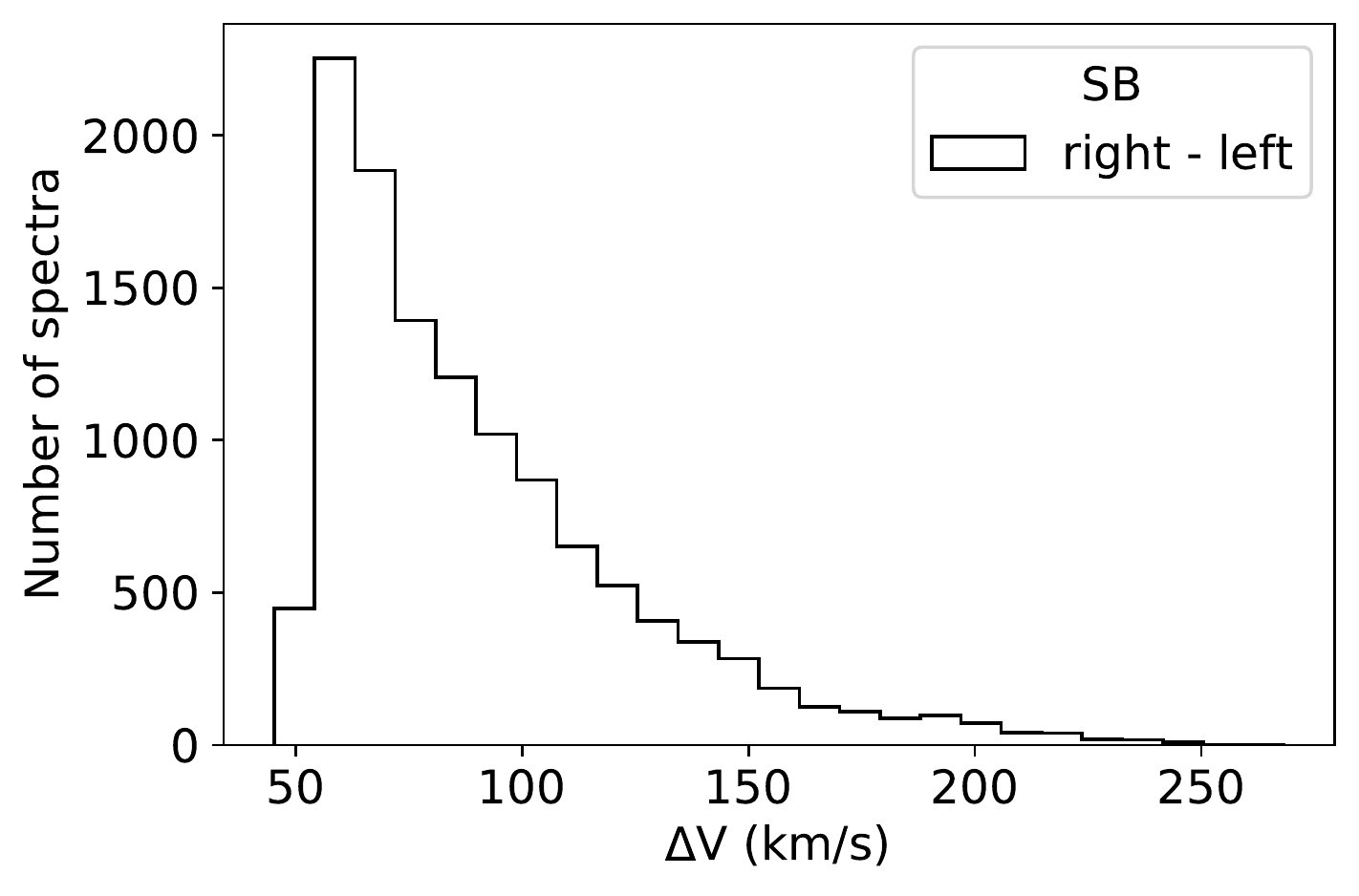}
    \includegraphics[width = 3in]{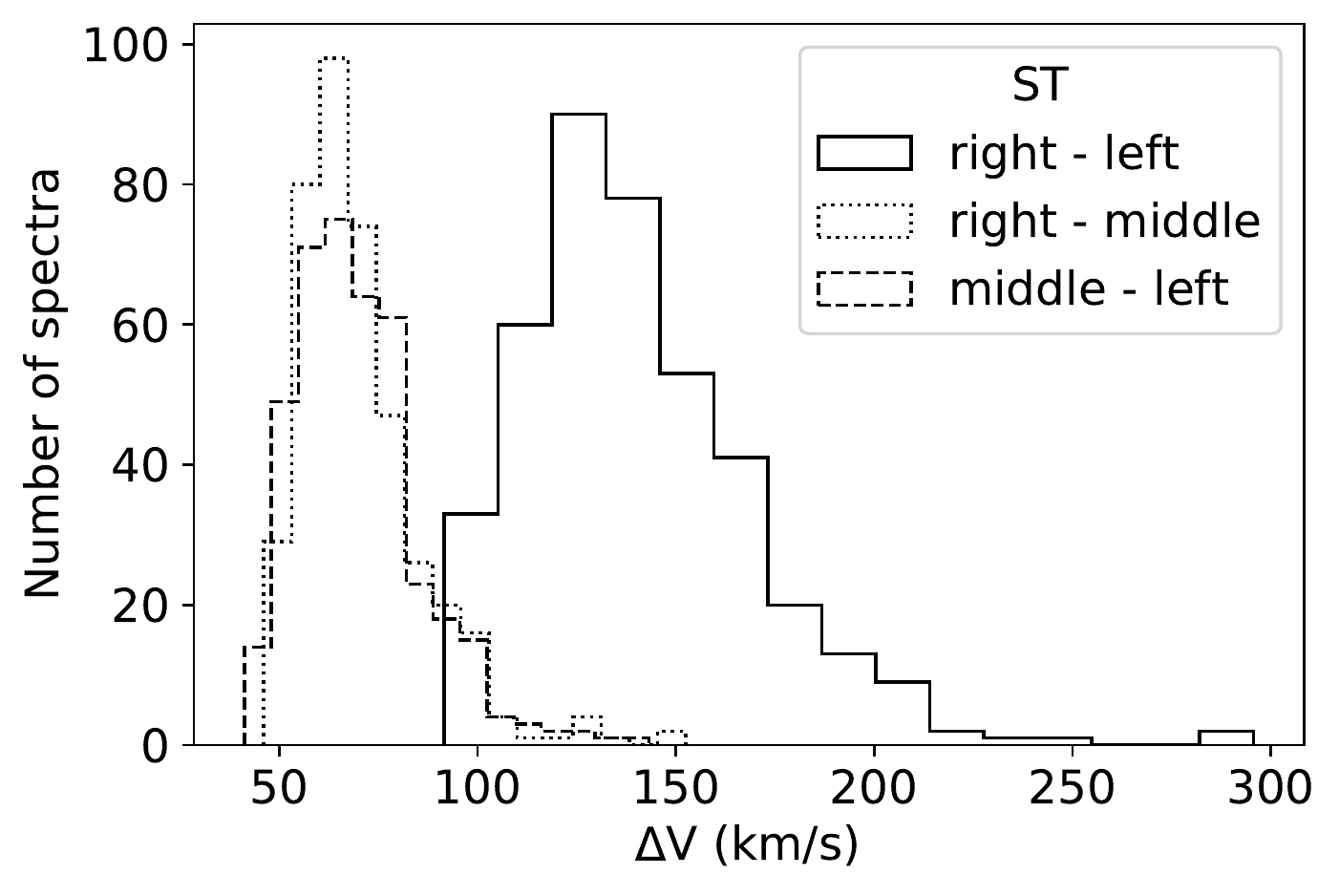}
    \caption{The number distribution of SB and ST spectra RV difference.}
    \label{fig: SB_dv}
\end{figure*}

A Monte-Carlo (MC) simulation is applied to estimate the uncertainties of our RV measurements. The main idea is to generate a set of simulated spectra based on each observed spectrum, and derive their RV values from the CCFs. For each RV component, we take the standard deviation of these RVs as its RV uncertainty.

For each observed spectrum, we generate 100 simulated spectra. For a simulated spectrum, the flux at each wavelength point is a random value generated from a Gaussian distribution, and the mean and variance of the Gaussian distribution are the flux and flux error of the corresponding observed spectrum. The CCFs of an observed spectrum and the 100 simulated spectra are presented in Figure\,\ref{fig: MC_CCF}. We notice that the amplitudes of the simulated CCFs are lower than that of the observed one, because the process of generating simulated spectra introduces additional error, therefore, the SNRs of the simulated spectra are lower than that of the observed spectrum.

It should be noted that if the number of RV components detected from the simulated spectrum is different from the observed one, this spectrum is discarded.

Figure\,\ref{fig: SNR_dv} shows the distribution of SNR and RV uncertainties of all the detected SB candidates. The scatter of the RV errors ($\sim$ 1\,km/s) is in line with the value of \cite{Liu2019} for LAMOST-MRS spectra.

\section{Results} \label{sec: result}

After examining the LAMOST-MRS spectra, we detect 3,133 SB and 132 ST candidates, which account for 1.2\% of the LAMOST-MRS stars.

\subsection{SB and ST candidates and RVs} \label{subsec: SB&ST}

There are 14,647 double- and 1,065 triple-line spectra having been detected with this method, and 11,648 double- and 388 triple-line spectra are confirmed after checking the CCF diagrams by human eyes. The proportions are 80\% and 36\% for the double- and triple-line spectra, respectively.

We present all the SB and ST candidates and their RV values and uncertainties in Table\,\ref{tab: SB_ST_rv}, including the celestial coordinates, \textit{source id} from \textit{Gaia} Early Data Release 3 \citep{Gaia_EDR3}, \textit{Gaia} G magnitude and the number of exposures. Four spectral information from the LAMOST data release: the LAMOST MRS plan name (planID), spectrograph ID (spID), fiber ID and Local Modified Julian Minute (LMJM), is also presented. It need to be pointed out that the RV values \textit{rv1}, \textit{rv2} and \textit{rv3} are arranged in order of values, as it is difficult to identify which RV component corresponds to each star when they have similar spectral type.

We consider a star to be ST candidate as long as one of its spectra has been detected three RV components. Accordingly, we classified 41 stars with both double- and triple-line spectra as ST candidates. The RV variations indicate that they are probably physical binaries or triples. Further checking the RV variations of these candidates will help us to identify more physical multiple star systems.

Figure\,\ref{fig: SB_ST_epoch} shows the distribution of the number of exposures of the detected candidates. There are 525 SB and 31 ST candidates with more than 6 observations, they are needed to be investigated as their orbital parameters may be derived by RV curves.

The detected RV difference limit is around 50\,km/s for all of the SB and ST candidates (Figure\,\ref{fig: SB_dv}), which meets the resolution of the LAMOST-MRS spectra. The largest RV differences are around 250\,km/s and 300\,km/s for SB and ST candidates, respectively.

In order to improve the detection efficiency, a human-free method is needed. Among the machine learning and deep learning techniques, the Recurrent Neural Network (RNN) is proved to be highly-performant for this classification \citep{2020arXiv200308618J}. To test the feasibility, we designed a RNN to derive the peak numbers from the CCF diagrams. Our RNN is trained by the CCF diagrams of the confirmed SB and ST candidates, and the details about the RNN can be found in Appendix\,\ref{App: detect_peaks}. The total classification accuracy of our RNN is 0.96, which is a significant improvement compared to the traditional method. Therefore, we will use RNN to detect SB and ST candidates in future study.

\subsection{New binary candidates} \label{subsec: new_SB}

We cross-match our LAMOST-MRS SB and ST candidates with these of the other binary catalogues, including $\rm S_{B^9}$ \citep{SB9}, the Geneva-Copenhagen Survey of Solar neighbourhood III \citep{GCS2009}, the RAVE SB2 \citep{RAVE_SB2}, the \textit{Gaia}-ESO multi-line SB \citep{Gaia-ESO_2017}, the Washington Visual Double Star \citep[WDS,][]{WDS}, the binaries of APOGEE \citep{APOGEE/IN-SYNC_2017,APOGEE_Bianry_2018}, the FGK binary stars of the GALAH survey \citep{GALAH_binary} and the third revision of Kepler Eclipsing Binary Catalog \citep[KEBIII,][]{KEBIII}, and there are 107 stars in common. The information about these stars are listed in Table\,\ref{tab: SB_ST_matches}.

Except for these common stars, there are 3,034 SB and 124 ST candidates (over 95\% of the total candidates) newly discovered. These RVs can be used to derive the orbital parameters of these systems \citep{PanYang, WangJX}.

\begin{longrotatetable}
\centering
\movetabledown=0.5in
\begin{deluxetable*}{ccrcccrrrccrrr}
    \centering
    \tablenum{1}
    \tabletypesize{\footnotesize}
    \tablecaption{The information of SB, ST candidates and their RVs.}
    \tablewidth{0pt}
    \label{tab: SB_ST_rv}
    \tablehead{
        \colhead{RA(2000)} & \colhead{DEC(2000)} & \colhead{Gaia source id} & \colhead{G(mag)} & \colhead{SB/ST} & \colhead{$N_{\rm Exp}$} & \colhead{planID} & \colhead{spID} & \colhead{fiberID} & \colhead{LMJM} & \colhead{SNR} & \colhead{rv1 (km/s)} & \colhead{rv2 (km/s)} & \colhead{rv3 (km/s)}
    }
    \startdata
    0.01948 & 57.48923 & 422589945455866496 & 12.1 & SB & 1 & HIP11784201 & 9 & 62 & 83604881 & 17 & $-36.7 \pm 0.7$ & $20.7 \pm 0.7$ & - \\
    0.13490 & 63.87285 & 431630508024107392 & 12.7 & SB & 1 & NGC778801 & 12 & 27 & 83650762 & 15 & $-57.0 \pm 3.3$ & $39.5 \pm 15.1$ & - \\
    0.23830 & 40.43928 & 2881984048448075264 & 12.4 & SB & 1 & HIP11776901 & 6 & 17 & 83609143 & 14 & $-58.5 \pm 0.8$ & $-6.9 \pm 0.5$ & - \\
    0.29125 & 58.69553 & 422768646150973696 & 12.2 & SB & 3 & NGC778901 & 12 & 240 & 83647859 & 15 & $-9.3 \pm 5.5$ & $62.6 \pm 1.6$ & - \\
    0.29125 & 58.69553 & 422768646150973696 & 12.2 & SB & 3 & NGC778901 & 12 & 240 & 83647873 & 14 & $-11.0 \pm 5.6$ & $63.7 \pm 7.8$ & - \\
    0.29125 & 58.69553 & 422768646150973696 & 12.2 & SB & 3 & NGC778901 & 12 & 240 & 83647886 & 16 & $-19.2 \pm 7.5$ & $59.8 \pm 8.5$ & - \\
    0.34767 & 41.11409 & 2882225867991453056 & 12.2 & SB & 9 & HIP11776901 & 6 & 39 & 83604805 & 21 & $-95.4 \pm 0.4$ & $32.4 \pm 0.7$ & - \\
    0.34767 & 41.11409 & 2882225867991453056 & 12.2 & SB & 9 & HIP11776901 & 6 & 39 & 83604818 & 20 & $-96.0 \pm 0.3$ & $31.0 \pm 0.8$ & - \\
    0.34767 & 41.11409 & 2882225867991453056 & 12.2 & SB & 9 & HIP11776901 & 6 & 39 & 83604832 & 22 & $-94.6 \pm 0.3$ & $30.9 \pm 0.7$ & - \\
    0.34767 & 41.11409 & 2882225867991453056 & 12.2 & SB & 9 & HIP11776901 & 6 & 39 & 83607704 & 27 & $-85.3 \pm 0.6$ & $9.2 \pm 0.4$ & - \\
    $\cdots$ & $\cdots$ & $\cdots$ & $\cdots$ & $\cdots$ & $\cdots$ & $\cdots$ & $\cdots$ & $\cdots$ & $\cdots$ & $\cdots$ & $\cdots$ & $\cdots$ & $\cdots$ \\
    \hline
    5.34918 & 58.28837 & 422132853553794816 & 12.9 & ST & 3 & NT002740N583314C02 & 3 & 187 & 84196452 & 49 & $-72.7 \pm 1.5$ & $-0.4 \pm 1.5$ & $61.1 \pm 2.2$ \\
    5.34918 & 58.28837 & 422132853553794816 & 12.9 & ST & 3 & NT002740N583314C02 & 3 & 187 & 84196476 & 47 & $-69.6 \pm 1.9$ & $-2.2 \pm 1.0$ & $58.6 \pm 2.0$ \\
    5.34918 & 58.28837 & 422132853553794816 & 12.9 & ST & 3 & NT002740N583314C02 & 3 & 187 & 84196499 & 49 & $-74.7 \pm 1.4$ & $-1.4 \pm 1.1$ & $63.1 \pm 2.4$ \\
    17.51845 & 2.00869 & 2538630790708445312 & 10.7 & ST & 4 & TD010605N031628K01 & 7 & 113 & 84118991 & 58 & $-20.9 \pm 0.5$ & $63.4 \pm 0.4$ & - \\
    17.51845 & 2.00869 & 2538630790708445312 & 10.7 & ST & 4 & TD010605N031628K01 & 7 & 113 & 84119014 & 50 & $-20.3 \pm 0.6$ & $64.6 \pm 0.5$ & - \\
    17.51845 & 2.00869 & 2538630790708445312 & 10.7 & ST & 4 & TD010605N031628K01 & 7 & 113 & 84119038 & 39 & $-18.5 \pm 0.6$ & $65.7 \pm 0.7$ & - \\
    17.51845 & 2.00869 & 2538630790708445312 & 10.7 & ST & 4 & TD010605N031628K01 & 7 & 113 & 84166383 & 46 & $-65.7 \pm 0.5$ & $-15.6 \pm 0.4$ & $79.5 \pm 0.5$ \\
    18.16407 & 45.78222 & 400867753213471744 & 11.9 & ST & 4 & TD012220N453143T01 & 14 & 68 & 84156353 & 15 & $-97.8 \pm 2.1$ & $0.2 \pm 3.1$ & - \\
    18.16407 & 45.78222 & 400867753213471744 & 11.9 & ST & 4 & TD012220N453143T01 & 14 & 68 & 84156377 & 14 & $-97.3 \pm 2.3$ & $-0.5 \pm 3.6$ & - \\
    18.16407 & 45.78222 & 400867753213471744 & 11.9 & ST & 4 & TD012220N453143T01 & 14 & 68 & 84156400 & 12 & $-95.3 \pm 1.6$ & $3.2 \pm 1.5$ & - \\
    18.16407 & 45.78222 & 400867753213471744 & 11.9 & ST & 4 & TD012220N453143T01 & 14 & 68 & 84156423 & 11 & $-96.3 \pm 2.2$ & $-46.4 \pm 3.5$ & $0.3 \pm 2.2$ \\
    $\cdots$ & $\cdots$ & $\cdots$ & $\cdots$ & $\cdots$ & $\cdots$ & $\cdots$ & $\cdots$ & $\cdots$ & $\cdots$ & $\cdots$ & $\cdots$ & $\cdots$ & $\cdots$ \\
    \enddata
    \tablecomments{The RV values \textit{rv1}, \textit{rv2} and \textit{rv3} are arranged in order of values. 41 ST candidates show both double- and triple-line spectra.}
\end{deluxetable*}
\end{longrotatetable}

\section{Discussion} \label{sec: discuss}

\subsection{Detection efficiency} \label{subsec: detect_eff}

To test the detection efficiency of the method, we use the radiative transfer code \textit{SPECTRUM} \citep{SPECTRUM} and the \textit{Kurucz} \citep{Kurucz} stellar atmosphere models to generate six synthetic spectra of given stellar parameters. They include cool and hot ($T_{\rm eff} = 5000, 8000$\,K), giant and dwarf ($\log{g}$ = 2.0, 4.0\,dex), metal-rich and metal-poor ([Fe/H] = 0.0, $-$2.0\,dex) models. All of the synthetic spectra have been reduced to the same wavelength range and resolution (R$\sim$7,500) as the LAMOST-MRS blue arm ones. We generate a set of double-line synthetic spectra by shifting a synthetic spectrum with different RVs and combining it with the original ones, and the synthetic double-line SBs have equal masses.

A similar MC simulation (see in Sect\,\ref{subsec: result_eRV}) is adopted to test the detection efficiency. We add the random Gaussian distribution noises to each point of the synthetic spectrum and generate 100 spectra for a specific SNR to detect RV components. We count the number of spectra detected as SBs, and the success rate represents the detection efficiency.

\begin{figure*}[]
    \plotone{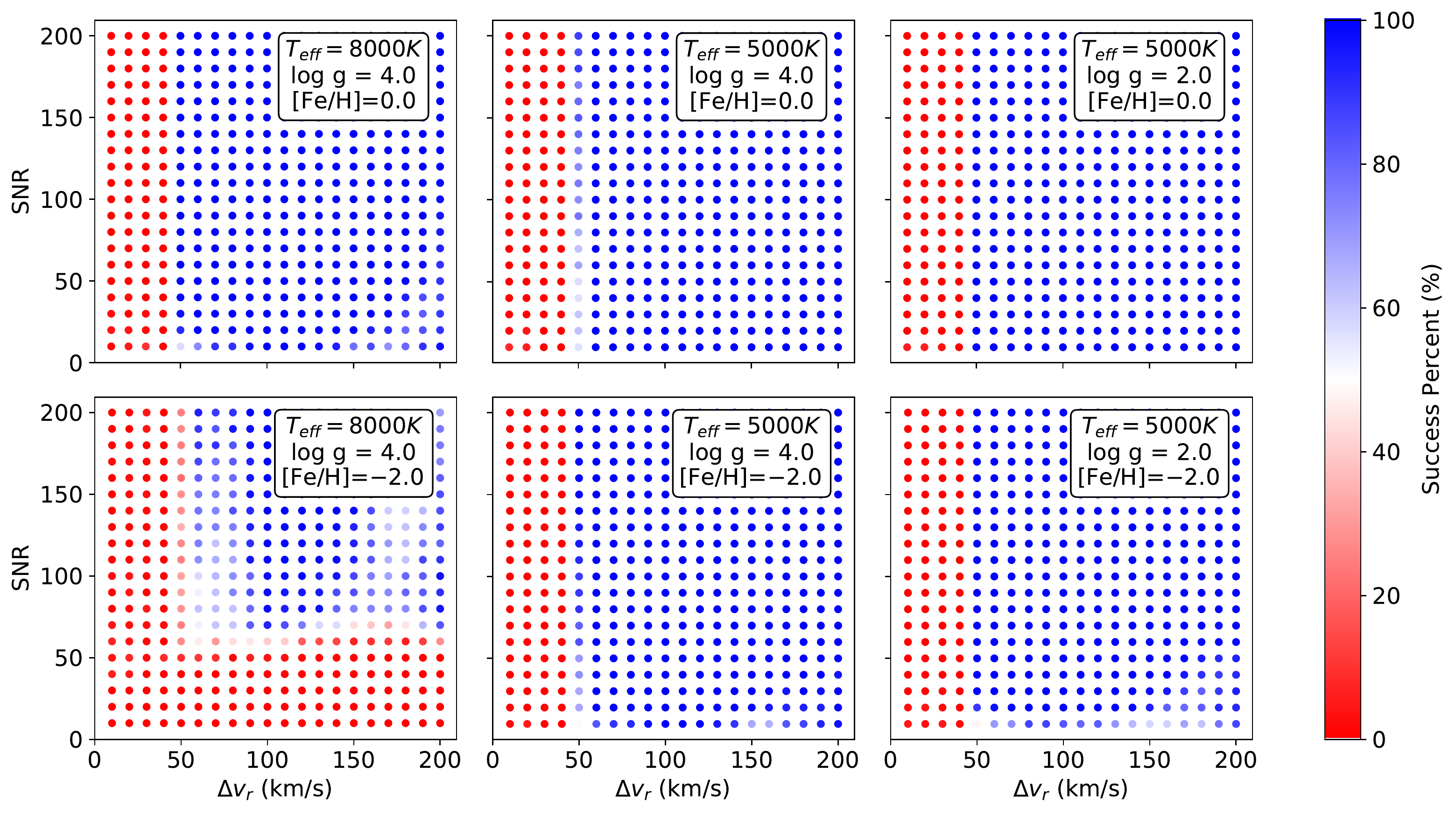}
    \caption{The detection efficiency of simulated SB in different atmospheric parameters, RV differences and SNR levels.}
    \label{fig: detect_eff} 
\end{figure*}

Our simulation results are presented in Figure\,\ref{fig: detect_eff}, and it can be seen that the detection limit of RV differences is between 40 and 50\,km/s, and this limit is consistent with the LAMOST-MRS resolution power. The detection efficiency of the method is similar except for hot metal-poor dwarfs.

\subsection{Stellar parameters} \label{subsec: par}

\begin{figure*}[]
    \plotone{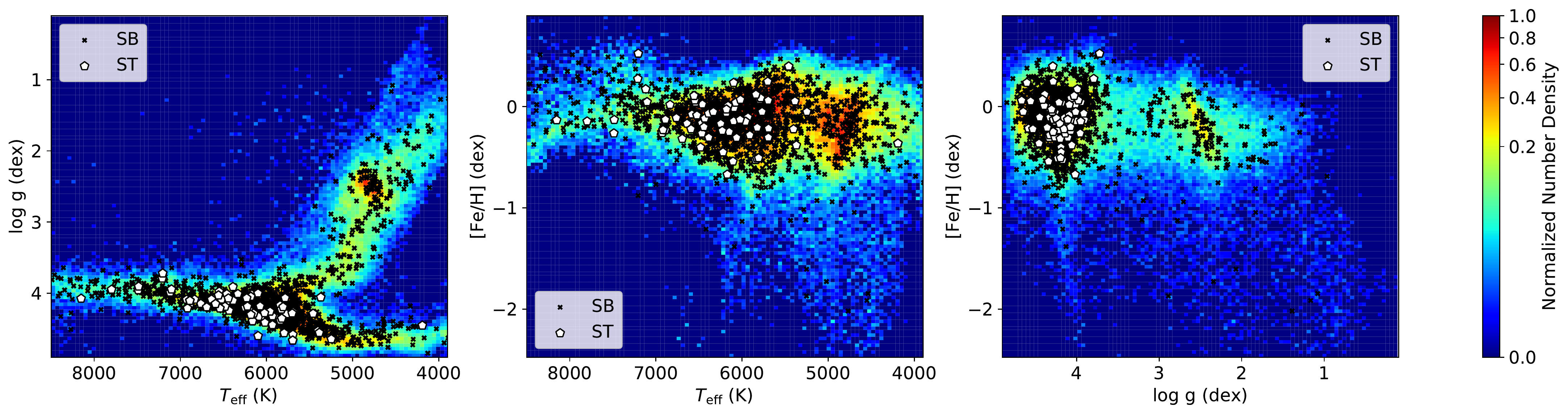}
    \caption{The distribution of stellar atmospheric parameters of common stars between the LAMOST DR7 Low-Resolution Spectroscopic Survey and the selected LAMOST-MRS data. The black cross are the SB candidates, and white pentagons are ST candidates. The left panel is $T_{\rm eff}$ \textit{vs} $\log{g}$ diagram, the middle panel is $T_{\rm eff}$ \textit{vs} [Fe/H] diagram, and the right panel is $\log{g}$ \textit{vs} [Fe/H] diagram. All the atmospheric parameters are provided by the LAMOST DR7 LRS.}
    \label{fig: par_SB_ST}
\end{figure*}

It is useful to investigate the distribution of stellar atmospheric parameters of SB and ST candidates, although the general method for determining the stellar atmospheric parameters may not be appropriate for multi-line spectra. We adopt the stellar atmospheric parameters of 1,470 SB and 65 ST candidates from the LAMOST DR7 low-resolution spectroscopic survey (LAMOST-LRS) \citep{LAMOST_DR1}, and present their distribution in $T_{\rm eff}$ \textit{vs} $\log{g}$, $T_{\rm eff}$ \textit{vs} [Fe/H], and $\log{g}$ \textit{vs} [Fe/H] in Figure\,\ref{fig: par_SB_ST}.

We note that about 90\% of the SB and all the ST candidates are on the main sequence, only 155 SBs may have one or two components on the red giant branch ($\log{g}<3.5$\,dex). The reason is that when the massive component climbs the red giant branch, its radius increases, and mass transfer occurs after it reaches the Roche lobe. The interaction between the components can modify the evolution and lead to fewer giant binary systems. The lack of metal-poor candidates is due to less low metallicity stars in our sample.

\subsection{Caveats} \label{subsec: Caveats}

There are some caveats we need to be pointed out before utilizing the SB or ST catalog for further studies. Since the completeness of the multi-line spectroscopic candidate catalog is not the primary goal of this work, the selection effects of the LAMOST-MRS data and the detecting process of the method must be considered in the further statistical studies.

Another caveat is that the SB and ST candidates are not necessarily physical binaries or triples. The CCFs of double- and triple-line spectra may be mimicked by emission lines or stellar pulsations, etc. \citep{Gaia-ESO_2017}. Considering the LAMOST fibers have a diameter of 3.3 arcsec \citep{LAMOST_2012}, one spectrum may contain light from multiple stars when they are very close to each other in the sky. Investigating the RV variations of the SB and ST candidates is essential to identify the physical binaries and triples.

\section{Conclusions} \label{sec: conclusion}

Based on the method of CCF and Gaussian smooth for the derivatives of CCF, we detected 14,647 double- and 1,065 triple-line spectra from the blue arm spectra with SNR$\geqslant$10 of the LAMOST DR7 MRS. After checking the CCF diagrams by human eyes, 11,648 double-line and 388 triple-line spectra are confirmed, the proportions are 80\% and 36\%, respectively. They belong to 3,133 SB and 132 ST candidates, these confirmed candidates account for 1.2\% of the LAMOST-MRS stars. Among the ST candidates, 41 of them show double- and triple-line cases, which indicates that they are physical binaries because of the variations of RVs. Comparing with the other binary star catalogs, we find that about 95 percent of them are newly discovered. For the 1,470 SB and 65 ST candidates, which have stellar atmospheric parameters from the LAMOST-LRS survey, 90\% of the SB and all the ST candidates are on the main sequence.

A Monte-Carlo simulation is used to estimate the RV uncertainties, and the RV error is of the order of 1\,km/s. Our results indicate that the detection limit of RV differences is between 40 and 50\,km/s.

The LAMOST-MRS survey is going on, and will provide a great opportunity of studying the binary and multiple star systems.

% \acknowledgments
\bigskip
\noindent \textbf{Acknowledgments} We thank the referee for the helpful comments which have helped us to improve the manuscript. Our research is supported by National Key R\&D Program of China No.2019YFA0405502, the National Natural Science Foundation of China under grant Nos. 12090040, 12090042, 12090044, 11833002, 11833006, 12022304, 11973052, 11973042 and U1931102. This work is supported by the Astronomical Big Data Joint Research Center, co-founded by the National Astronomical Observatories, Chinese Academy of Sciences and Alibaba Cloud. H.-L.Y. acknowledges the supports from Youth Innovation Promotion Association, Chinese Academy of Sciences (id. 2019060) and NAOC Nebula Talents Program. Guoshoujing Telescope (the Large Sky Area Multi-Object Fiber Spectroscopic Telescope LAMOST) is a National Major Scientific Project built by the Chinese Academy of Sciences. Funding for the project has been provided by the National Development and Reform Commission. LAMOST is operated and managed by the National Astronomical Observatories, Chinese Academy of Sciences.

\bibliographystyle{aasjournal}
\bibliography{sample63}

\appendix
\setcounter{figure}{0}

\section{Detecting Peaks} \label{App: detect_peaks}
\renewcommand{\thefigure}{A\arabic{figure}}

We design a simple Recurrent Neural Network (RNN) to inspect the peak numbers of CCF diagrams automatically. Because of the incremental data of the LAMOST-MRS survey, a human-free classification method is urgently need to analyze these spectra. Among mostly-used machine learning and deep learning (DL) techniques, RNN proves to be highly-performant for the classification of 1-dimension data \citep{2020arXiv200308618J} compared to the Convolutional Neural Network (CNN).

Recurrent neural network (RNN) refers to a class of artificial neural networks where network architecture composed of interconnected nodes through a directed graph along a temporal sequence. The Long-Short Term Memory \citep[LSTM,][]{hochreiter1997long} is one of the variants of RNN with gated state or memory. The LSTM has been found extremely successful in many applications, such as speech recognition \citep{graves2013speech}, handwriting generation \citep{graves2013generating}, and machine translation \citep{sutskever2014sequence}. At the same time, RNNs have been applied on astronomical data \citep{2020arXiv200308618J} including variable stars \citep{2018NatAs...2..151N}, periodic variable stars \citep{2019ApJ...877L..14T} and supernovae classifications, \citep{2017ApJ...837L..28C} and online transient event detection \citep{2019PASP..131k8002M, 2020MNRAS.491.4277M}. 

\begin{figure*}[h!]
    \centering
    \includegraphics[width = 3in]{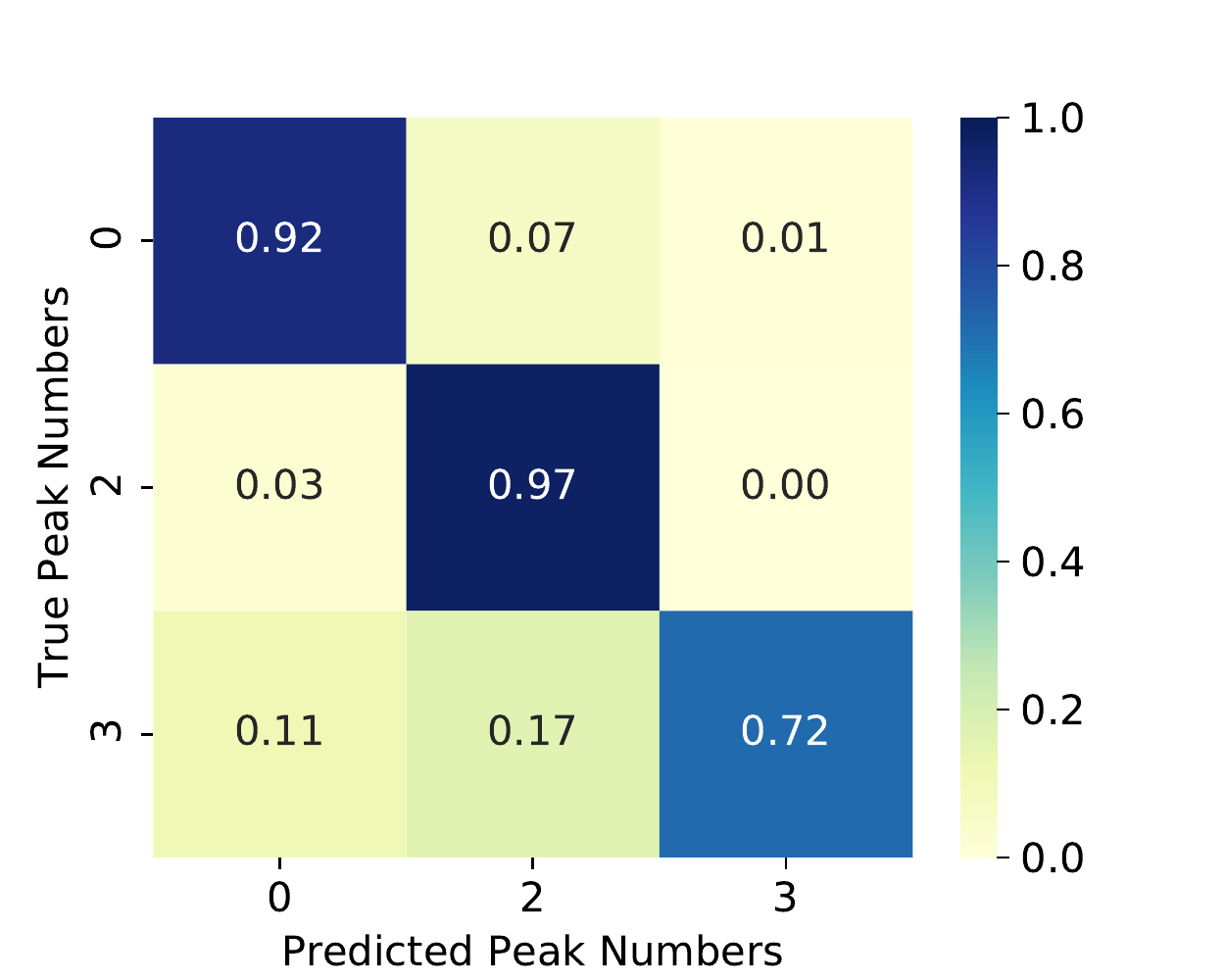}
    \caption{Confusion matrix obtained on the Test set predictions for the RNN classifier. The values in each box refer to the number of predictions versus the true labels.}
    \label{fig: confusion_mat}
\end{figure*}

We set up our RNN with a 3-layer LSTM, the hidden size of 256 units, and 3 output features. This RNN is trained by 12,136 CCF diagrams with labels as number of peaks (0, 2, 3) together with 4,046 groups of test data. We split the training data into batches to speed up the training process. After 100 training epochs, we stop the training process to avoid the overfitting (with a training loss of 0.003 and test loss of 0.100). The final classification accuracy and precision are shown in Table\,\ref{tab: classification} and the confusion matrix obtained on the Test set predictions for the RNN classifier is shown in Figure\,\ref{fig: confusion_mat}. 

\begin{table}[!h]
    \label{tab: classification}
    \tablenum{A1}
    \begin{center}
        \caption{The precision, recall and f1-score of the prediction of classification.}
        \begin{tabular}{cccc}
        \hline
        \hline
        Peak number & Precision & Recall & F1-score\\
        \hline
        0 & 0.91 & 0.92 & 0.92\\
        2 & 0.97 & 0.97 & 0.97\\
        3 & 0.84 & 0.72 & 0.78\\
        \hline
        \end{tabular}
    \end{center}
\end{table}

F1-score is the harmonic mean of precision and recall:

\begin{equation}
    F_1 = \frac{2}{{\rm precision}^{-1} + {\rm recall}^{-1}}
\end{equation}

The recognization of two peaks is performed well (precision $\sim 0.97$) since there are the most training data of two peaks CCFs. On the contrary, the group of three peaks is the most difficult one for recognization due to less data. Totally, the final classification accuracy is 0.96, which means that, for a given CCF diagram, we can detect the correct peak number successfully by the probability of 0.96. This method can improve the productiveness and robustness. We use the deep learning method along with human eyes to confirm our results.

\clearpage
\section{Common SB and ST candidates with other binary catalogs} \label{App: common_stars}

\startlongtable
\begin{deluxetable*}{rrrccrr}
    \tablenum{B1}
    \tabletypesize{\small}
    \tablecaption{List of the common SB and ST candidates with other binary catalogs.}
    \tablewidth{0pt}
    \label{tab: SB_ST_matches}
    \tablehead{
        \colhead{RA(2000)} & \colhead{DEC(2000)} & \colhead{Gaia source id} & \colhead{G} & \colhead{SB/ST} & \colhead{$N_{\rm Exp}$} & \colhead{Catalog} 
    }
    \startdata
    4.31148 & 0.55870 & 2545576336943059072 & 12.7 & SB & 3 & GALAH\\
    6.35503 & 59.11604 & 428267308105062528 & 11.1 & SB & 4 & WDS\\
    9.78481 & 59.81764 & 425556805786609408 & 10.9 & SB & 14 & WDS\\
    10.17108 & 57.76887 & 424898679355325312 & 11.3 & SB & 2 & WDS\\
    13.49841 & 10.08698 & 2582339740871439360 & 12.2 & SB & 1 & GALAH\\
    15.04087 & 12.21674 & 2584289591599704064 & 11.5 & SB & 1 & GALAH\\
    15.23534 & 11.58956 & 2584040273042478208 & 12.8 & SB & 2 & GALAH\\
    19.44386 & 0.62884 & 2535048989846592768 & 13.1 & SB & 2 & GALAH\\
    29.76969 & 58.39781 & 505491850882835584 & 11.0 & ST & 3 & WDS\\
    31.96472 & 13.79138 & 77202449462513024 & 12.1 & SB & 3 & APOGEE\\
    46.84714 & 40.80111 & 239854583246385408 & 10.3 & SB & 1 & WDS\\
    46.91366 & 66.17673 & 492339320985355648 & 10.6 & SB & 3 & WDS\\
    50.59945 & 51.66556 & 442912798683449088 & 12.1 & ST & 9 & WDS\\
    51.47665 & 18.13560 & 55926869402261760 & 12.5 & SB & 4 & GALAH\\
    51.85738 & 19.01199 & 57544044849005440 & 13.0 & SB & 2 & GALAH\\
    52.37579 & 19.41492 & 57640320835799808 & 12.3 & SB & 4 & GALAH\\
    55.16722 & 45.77760 & 244833962172857728 & 11.2 & ST & 3 & WDS\\
    62.05420 & 19.94418 & 51832421242688128 & 12.4 & SB & 3 & WDS\\
    63.01352 & 48.26905 & 246251095218220160 & 11.0 & SB & 1 & WDS\\
    63.01540 & 21.94700 & 52684199156684032 & 8.9 & SB & 3 & SB9, WDS\\
    63.29504 & 61.55950 & 475258132965412224 & 13.1 & SB & 3 & WDS\\
    66.44780 & 18.31499 & 3314466360138302592 & 13.0 & SB & 2 & GALAH\\
    68.20363 & 16.02178 & 3312652582565893632 & 13.0 & SB & 9 & GALAH\\
    71.04484 & 26.76769 & 154533339923810688 & 13.7 & SB & 11 & GALAH\\
    71.43970 & 23.53370 & 146573528573114240 & 12.9 & SB & 4 & GALAH\\
    71.99216 & 22.95037 & 3413141584496385792 & 12.0 & SB & 11 & GALAH\\
    72.16204 & 15.36070 & 3405024886580964736 & 12.9 & SB & 3 & GALAH\\
    72.21994 & 14.71636 & 3308722103373770496 & 13.4 & SB & 3 & GALAH\\
    72.40034 & 56.28887 & 274601899464974592 & 10.6 & SB & 3 & WDS\\
    73.58480 & 20.85980 & 3411716548707061376 & 11.7 & ST & 8 & GALAH\\
    73.72125 & 24.23559 & 3419659356282987008 & 12.5 & SB & 9 & GALAH\\
    74.90812 & 21.79690 & 3412251701632573312 & 11.7 & ST & 5 & GALAH\\
    75.05998 & 24.13008 & 3419410763576132352 & 10.9 & ST & 12 & WDS\\
    76.05618 & 24.30754 & 3418774008905352704 & 10.3 & SB & 3 & WDS\\
    77.57841 & 24.48785 & 3419012499849582464 & 11.5 & SB & 3 & GALAH\\
    83.88432 & 35.85607 & 183166645640035584 & 11.9 & SB & 3 & WDS\\
    92.23265 & 22.14776 & 3423547328182688640 & 10.9 & SB & 1 & WDS\\
    98.13701 & 8.59674 & 3326258557925415936 & 11.2 & SB & 3 & WDS\\
    98.72293 & 9.43785 & 3326828345466157952 & 10.8 & SB & 5 & WDS\\
    100.44896 & 9.86729 & 3326737665818107008 & 13.1 & SB & 2 & Gaia ESO\\
    101.89452 & 8.47141 & 3134190163070718464 & 10.4 & ST & 7 & WDS\\
    101.93376 & 23.12819 & 3379557723381624832 & 11.2 & SB & 3 & WDS\\
    102.21957 & 17.33513 & 3358396660030939776 & 12.0 & SB & 6 & WDS\\
    103.79022 & 22.80749 & 3380050961721719680 & 10.5 & SB & 7 & WDS\\
    106.83926 & 45.75925 & 977374646347674624 & 12.5 & SB & 3 & WDS\\
    115.99661 & 24.74451 & 867894834757702400 & 13.5 & SB & 1 & WDS\\
    123.65855 & 14.73966 & 655293644368357120 & 11.7 & SB & 3 & GALAH\\
    123.72821 & 16.71239 & 655829140890694016 & 13.0 & SB & 5 & GALAH\\
    123.85360 & 16.22931 & 655726439632891648 & 12.7 & SB & 3 & GALAH\\
    124.00038 & 17.45520 & 657004106505230336 & 12.6 & SB & 4 & GALAH\\
    124.14557 & 19.41296 & 663514555370788224 & 12.4 & SB & 7 & GALAH\\
    124.24190 & 17.65389 & 657014483146180352 & 11.8 & SB & 4 & GALAH\\
    124.47248 & 14.38159 & 652210957361785728 & 11.7 & SB & 1 & GALAH\\
    124.72561 & 18.90909 & 663247374044278912 & 11.8 & SB & 5 & GALAH\\
    124.96669 & 17.39997 & 656285232057964800 & 13.7 & SB & 11 & GALAH\\
    126.15704 & 16.84889 & 656018841006561792 & 11.7 & SB & 3 & GALAH\\
    126.95277 & 13.03253 & 650982218756793984 & 12.8 & SB & 5 & GALAH\\
    126.95640 & 11.58378 & 601155100565167872 & 12.9 & SB & 3 & GALAH\\
    127.19336 & 19.00738 & 662878900210497408 & 11.5 & SB & 11 & GALAH\\
    127.23194 & 18.76372 & 662309731144349184 & 10.9 & SB & 5 & GALAH\\
    130.17758 & 17.23037 & 658488928236761600 & 13.8 & SB & 2 & GALAH\\
    130.86018 & 12.41902 & 602244372990566400 & 11.1 & SB & 3 & WDS\\
    131.03050 & 20.07688 & 661380815618317312 & 10.0 & SB & 3 & SB9\\
    132.52700 & 23.75079 & 689509857812011392 & 12.4 & SB & 11 & WDS\\
    132.70634 & 12.28767 & 605014833054650880 & 11.5 & SB & 3 & APOGEE\\
    132.82328 & 11.66001 & 604897803786032768 & 13.4 & SB & 11 & GALAH\\
    132.85539 & 12.04901 & 604972089540120832 & 13.4 & SB & 2 & APOGEE, GALAH\\
    133.01907 & 19.60290 & 660730076531974016 & 11.2 & SB & 14 & GALAH\\
    133.27465 & 10.34295 & 597812482136772608 & 12.8 & SB & 24 & GALAH\\
    133.34198 & 19.34574 & 660661013457939200 & 11.6 & SB & 2 & GALAH\\
    133.41641 & 23.21572 & 689235946273862016 & 13.1 & SB & 21 & GALAH\\
    133.57891 & 12.65805 & 605136088570783104 & 11.2 & SB & 13 & APOGEE, WDS\\
    133.68605 & 11.50144 & 604716006409635456 & 10.3 & SB & 11 & APOGEE, GALAH\\
    133.89588 & 16.75347 & 611576370555983872 & 12.3 & SB & 10 & GALAH\\
    133.99515 & 22.92465 & 686164387525931648 & 13.0 & SB & 29 & GALAH\\
    143.96777 & 37.43883 & 799090386388958720 & 13.3 & SB & 6 & APOGEE\\
    144.06575 & 37.52891 & 799092516692729344 & 10.3 & SB & 5 & SB9\\
    168.54278 & 4.98812 & 3816819414549676544 & 11.1 & SB & 3 & GALAH\\
    170.48610 & 3.71580 & 3812680406106190848 & 12.7 & SB & 3 & WDS\\
    173.17225 & 1.59537 & 3799929369758876032 & 13.4 & SB & 6 & GALAH\\
    173.69728 & 2.06516 & 3800061409937591552 & 12.7 & SB & 1 & GALAH\\
    174.26187 & 2.27368 & 3799346834754684544 & 13.3 & SB & 3 & GALAH\\
    179.39510 & 3.22525 & 3893093635680954496 & 10.7 & SB & 3 & GALAH\\
    184.90873 & 25.29859 & 4008485212056046208 & 11.3 & SB & 8 & APOGEE\\
    185.32486 & 50.97667 & 1547679958899540224 & 10.4 & SB & 3 & WDS\\
    207.00384 & -0.78145 & 3662029274238005120 & 11.5 & SB & 3 & WDS\\
    207.24577 & -8.31973 & 3619010057167751808 & 10.9 & SB & 10 & GALAH\\
    207.60620 & -6.40342 & 3620621460177550592 & 11.2 & SB & 5 & GALAH\\
    226.29512 & 26.56544 & 1268407888092649344 & 10.5 & SB & 1 & APOGEE\\
    263.88506 & 23.17511 & 4557402515188596224 & 10.9 & SB & 1 & SB9, WDS\\
    266.19860 & 6.03134 & 4474130344924152192 & 12.1 & SB & 2 & GALAH\\
    276.72032 & 25.83382 & 4584847493651016064 & 12.1 & SB & 2 & WDS\\
    284.75986 & 41.04340 & 2104025523931432704 & 10.5 & SB & 2 & KEBIII\\
    289.86104 & 41.40818 & 2101467956809222912 & 10.5 & SB & 1 & KEBIII\\
    289.88398 & 45.30403 & 2127090146848465152 & 11.6 & SB & 3 & KEBIII\\
    290.48320 & 43.62231 & 2126083307735831936 & 12.3 & SB & 1 & KEBIII\\
    290.51746 & 40.69681 & 2101192082470870912 & 11.3 & SB & 3 & KEBIII\\
    290.80668 & 42.34393 & 2101942327356009216 & 12.0 & SB & 3 & APOGEE, WDS, KEBIII\\
    290.95133 & 40.54793 & 2101510803402761344 & 14.1 & SB & 22 & KEBIII\\
    291.31991 & 43.59553 & 2126044240713250304 & 10.7 & SB & 26 & KEBIII\\
    292.05871 & 43.92521 & 2125971226269667584 & 12.7 & SB & 6 & KEBIII\\
    292.29674 & 44.28051 & 2126356983051726720 & 13.2 & SB & 25 & KEBIII\\
    292.71791 & 41.92241 & 2077667962475652864 & 10.1 & ST & 1 & KEBIII\\
    293.20532 & 42.19786 & 2077683321279046016 & 12.1 & SB & 2 & APOGEE\\
    294.47828 & 46.86376 & 2128532603025164288 & 12.5 & SB & 3 & KEBIII\\
    305.54277 & 40.21833 & 2067518679871015168 & 11.7 & SB & 4 & WDS\\
    332.42002 & 58.31327 & 2199510343510448768 & 10.3 & SB & 3 & WDS\\
    \enddata
\end{deluxetable*}

\,

\end{document}